\documentclass[useAMS,usenatbib]{mnras}
\usepackage[pdftex]{graphicx} 
\usepackage{hyperref}
\usepackage{xcolor} 
\usepackage{lscape}
\usepackage{calc}
\usepackage{amsmath}
\usepackage{afterpage}
\usepackage{bm}
\usepackage{amssymb, physics}
\defcitealias{Jespersen2025a}{J25}

\hypersetup{colorlinks,linkcolor={red!50!black}, citecolor={blue!50!black},urlcolor={blue!80!black} } 

\title[Conformity of high-$z$ massive galaxies]
{Think inside the box: cosmic variance and large-scale conformity of high-redshift massive galaxies in the FLAMINGO simulations}

\author[Lim et al.]
{Seunghwan Lim$^{1,2}$\thanks{E-mail: sl2207@cam.ac.uk},
Sandro Tacchella$^{1,2}$, 
Roberto Maiolino$^{1,2}$, 
Christopher C. Lovell$^{1}$, 
Joop Schaye$^{3}$
\\
\vspace*{6pt} \\
$^{1}$Kavli Institute for Cosmology, University of Cambridge, Madingley
Road, Cambridge, CB3 0HA, UK \\
$^{2}$Cavendish Laboratory, University of Cambridge, 19 JJ Thomson
Avenue, Cambridge, CB3 0HE, UK \\
$^{3}$Leiden Observatory, Leiden University, PO Box 9513, 2300 RA Leiden, the Netherlands 
\
}

\begin{document} 

\pagerange{\pageref{firstpage}--\pageref{lastpage}}

\date{\today}
\pubyear{2025}

\maketitle

\label{firstpage}

\begin{abstract}
We use the highest-resolution FLAMINGO hydrodynamical simulation to quantify cosmic variance and large-scale coherence in the evolution of massive galaxies at high redshift. FLAMINGO combines a $(1\,\mathrm{cGpc})^3$ volume with baryonic resolution sufficient to identify ${\gtrsim}\,10^3$ independent JWST-like survey volumes of $(100\,\mathrm{cMpc})^3$, providing unprecedented statistics to characterize the extremes of cosmic variance. At $z\,{\simeq}\,6$, the total variance in the number of haloes with $M_{200}\,{\simeq}\,10^{11.5}\,\mathrm{M_\odot}$ (or $M_\ast\,{\simeq}\,10^{10}\,\mathrm{M_\odot}$) is 2--3 times the Poisson expectation, while this ratio decreases with redshift. Similarly, at $z\,{\gtrsim}\,4$, the variance in the most massive halo per JWST-like field is twice the Poisson prediction. We find a pronounced large-scale \emph{conformity}: in volumes ranked by the stellar mass of their most massive galaxy ($M_{\ast,\mathrm{max}}$), the stellar-to-halo mass relation and star-formation efficiency are coherently elevated or suppressed throughout the full $(100\,\mathrm{cMpc})^3$ volume. When accounting for galaxies outside the volume, this signal persists only to radii $\lesssim 50\,\mathrm{cMpc}$, demonstrating that the detectable conformity is enhanced by the survey footprint. Moreover, $M_{\ast,\mathrm{max}}$ is a better predictor of the volume-wide efficiency of massive galaxies than the total number counts, which mainly trace clustering. Finally, the stellar fraction of the most massive galaxies peaks at $f_\ast\,{=}\,M_\ast\,/\,(M_{200}f_{\rm b,cosmic})\,{\simeq}\,0.2$ at $z\,{\simeq}\,5$, with a narrower dispersion in $f_\ast$ at fixed redshift and stronger redshift evolution than commonly assumed. These results show that both cosmic variance and footprint-confined conformity must be modelled when interpreting early massive galaxy populations in JWST fields. 
\end{abstract} 


\section[intro]{Introduction}
\label{sec_intro}

The masses and abundance of some early massive galaxies observed with JWST have been claimed to be in tension with the $\Lambda$CDM model or at least with the widely accepted galaxy formation paradigm \citep[e.g.,][]{Boylan-Kolchin2023, Labbe2023}. The claim is that these galaxies are too massive and too numerous to exist at such early cosmic times given the age of the Universe. It is important to emphasize, however, that this claim is contingent on stellar mass estimates and the mapping to halo mass, which are themselves subject to significant uncertainties. Indeed, some state-of-the-art cosmological simulations, such as COLIBRE \citep{Schaye2025, Chaikin2025a}, demonstrate that the observed evolution of the stellar mass function and the abundance of quenched galaxies, including constraints from JWST, are consistent with the standard $\Lambda$CDM framework \citep{Chaikin2025b}. The observational evidence fueling this debate stems from two primary fronts. Studies have focused on massive quiescent galaxies at $3 < z < 5$, whose formation times can be as early as $z\gtrsim8$ \citep[e.g.,][]{Carnall2023, Carnall2024, Glazebrook2024, Nanayakkara2024, Baker2025, Turner2025}. Meanwhile, other surveys have revealed similar populations directly at ultra-high redshifts \citep[e.g.,][]{Akins2023, Weibel2024, Harvey2025, Shuntov2025}. Should the tension persist, proposed solutions that would modify the standard galaxy formation model include a top-heavy initial mass function \citep[IMF; e.g.,][]{Inayoshi2022, vanDokkumConroy2024}, an elevated star-formation efficiency \citep[SFE; e.g.,][]{Dekel2023, Li2024}, and UV variability arising from stochastic star formation \citep[e.g.,][]{Shen2023, Sun2023, KravtsovBelokurov2024}. 

Another possible resolution, however, that has been paid less attention to, is the role of cosmic variance (CV) in these discoveries. Namely, observations may have preferentially sampled regions of enhanced clustering and accelerated growth, where large-scale density fluctuations produce biased populations of extreme objects \citep[e.g.,][]{Moster2011, Steinhardt2021, Lim2024, Jespersen2025a}. JWST and other high-redshift surveys are particularly vulnerable to this effect, given the limited volumes they probe. A key challenge in quantifying CV is the lack of robust tools. Most high-redshift studies rely on the CV calculator of \citet{Moster2011} \citep[e.g.,][]{Valentino2023, Navarro-Carrera2024, Weibel2024, Harvey2025}, while others \citep[e.g.,][]{Finkelstein2023, Yung2024} have used the luminosity-based calculator of \citet{Bhowmick2020} from the BlueTides simulation \citep{Feng2016}. However, the applicability of these tools to the early Universe is uncertain. The \citet{Moster2011} model is based on a $(100\,\mathrm{cMpc})^3$ simulation and calibrated at $z\,{\simeq}\,0$, and the BlueTides-based model, while designed for high redshift, is also limited by its $(400\,\mathrm{cMpc}/h)^3$ volume. A more fundamental limitation of the \citet{Moster2011} approach is that it ignored the scatter in the stellar-to-halo mass relation (SMHMR), an effect that itself is a significant source of CV \citep{Jespersen2025a}. As another approach to evaluate the field-to-field variation of the high-$z$ galaxy properties and interpret observations of only a few rare, extreme systems, \citet{Lovell2023} adopted the framework of the Extreme Value Statistics (EVS; \citealt{Gumbel1958, KotzNadarajah2000}). However, their method accounted only for Poisson noise in sample statistics, neglecting the additional variance from large-scale clustering, i.e. CV. 

\citet{Jespersen2025a, Jespersen2025b} significantly advanced previous work by calibrating their analysis to the UniverseMachine \citep{Behroozi2019}, which enabled a robust incorporation of CV in galaxy abundance and mass at ultra-high redshifts. Their analysis reveals that the total variance, including CV, exceeds the predictions from Extreme Value Statistics (EVS) by \citet{Lovell2023} at $z\,{\gtrsim}\,9$ for small, JWST pencil-beam surveys. This effect is significantly reduced for wider-area surveys like COSMOS-Web due to their larger sampling volume. A key insight from \citet[][J25 hereafter]{Jespersen2025a} is that CV impacts number counts and the most massive expected galaxies ($M_{\rm max}$) in fundamentally different ways. Contrary to a naive expectation, \citetalias{Jespersen2025a} found that increased CV does not enhance the positive skewness of the $M_{\rm max}$ distribution but instead extends the positive tail of the number count distribution. This implies that cosmic variance could resolve the ``too many'' problem for early galaxies but fails to alleviate the tension posed by overly massive systems. This finding is based on 32 independent UniverseMachine lightcones, with fields spanning from 17$\arcmin\times41\arcmin$ (mimicking COSMOS) to 69$\arcmin\times32\arcmin$ (GOODS-N). \citetalias{Jespersen2025a} sampled these lightcones in redshift slices of $\Delta z\,{=}\,0.5$, corresponding to volumes of 0.78 to 2.5$\times 10^6$\,cMpc$^{3}$ at $z\,{=}\,7$--7.5. However, as the authors note, this approach uses fixed realizations of the CV and does not account for variance on larger scales. \citetalias{Jespersen2025a} find that this ``variance of the cosmic variance" can be as large as an order of magnitude, and its inclusion further suppresses the overall $M_{\rm max}$ distribution, thereby exacerbating the tension for early massive galaxies.

These findings from \citetalias{Jespersen2025a} collectively demonstrate that the cosmic variance in the total mass, which is dominated by dark matter, cannot resolve the puzzle of overly massive early galaxies. Consequently, the only remaining avenue to address this tension lies in the scatter in baryonic physics, as reflected e.g. in the SMHMR. Indeed, \citetalias{Jespersen2025a} explored this possibility by introducing a random scatter into the SMHMR within their model, finding that it significantly broadens the expected range of $M_{\rm max}$, including its positive-tail extremes. However, their test assumed a purely stochastic scatter, ignoring any environmental dependence as the physical driver of deviations from the mean SMHMR. A critical unanswered question is whether this scatter exhibits large-scale coherence, namely whether the star-formation efficiency and evolution of massive galaxies at $z\,{\gtrsim}\,7$ display conformity on scales exceeding those of protoclusters (${\gtrsim}\,10$\,cMpc). This concept, termed large-scale galactic conformity, extends the established phenomenon of galaxy conformity from low redshift, where it is traced by correlations in quenching, color, and other properties with environment \citep[e.g.,][]{Weinmann2006, Knobel2015, Paranjape2015, Lim2016}, to the epoch of reionization and to scales of up to 100\,cMpc. 

On protocluster scales of ${\simeq}$\,10\,cMpc \citep[e.g.,][]{Chiang2013, Chiang2017, Lovell2018}, observational and theoretical studies suggest that galaxies in overdense environments at $z\,{\gtrsim}\,5$ may experience accelerated growth and enhanced star formation compared to their field counterparts, pointing to environmentally-driven baryonic processes \citep[e.g.,][]{Lim2021, Lim2024, Helton2024b, Baxter2025, Fudamoto2025, Morokuma-Matsui2025}. However, whether this environmental conformity extends to scales significantly larger than individual protoclusters remains an open question. If such large-scale coherence exists, it would indicate systematic variations in galaxy formation efficiency across the cosmic web, rather than mere stochastic scatter. This would provide a compelling, physically-motivated mechanism to explain the positive outliers in the stellar mass-halo mass relation and potentially resolve the tension posed by the abundance of overly massive galaxies at ultra-high redshift. 

In this work, we quantify the impact of cosmic variance on both the number counts and masses of high-redshift massive galaxies using the FLAMINGO cosmological simulation \citep{Schaye2023, Kugel2023}. As one of the few full-hydro simulations with a sufficiently large volume, FLAMINGO enables robust probing of the CV across JWST-typical (100\,cMpc)$^3$ fields with statistical significance up to 3\,$\sigma$, making it uniquely suited for this investigation. We also explore the star-formation efficiency and baryon content of the massive galaxies (of $M_\ast\,{\gtrsim}\,10^{10}\,{\rm M}_\odot$ at $z\,{\simeq}\,7$), and how they evolve across cosmic time. Our analysis suggests that there can be ${\simeq}\,100$\,cMpc-scale coherence in the evolution of massive galaxies at $z\,{\gtrsim}\,5$. We also demonstrate that the mass of the most massive galaxy is a unique environmental indicator of a given survey volume, providing new insights into how extreme objects may reflect broader cosmic conditions.

This paper is structured as follows. Sect.~\ref{sec_method} describes the FLAMINGO simulation and our methods. We present our main results in Sect.~\ref{sec_result}. In Sect.~\ref{sec_discussion}, we discuss our results in more detail and their observational implications. We summarize out findings in Sect.~\ref{sec_summary}.

\section[methods]{methods}
\label{sec_method}

\subsection{The FLAMINGO simulation}
\label{ssec_sample}

We adopt the data from the FLAMINGO simulation \citep{Schaye2023, Kugel2023}. As noted in Sect.~\ref{sec_intro}, FLAMINGO is one of a few hydrodynamical simulations that achieves both the box size and mass resolution of the baryonic elements suitable for probing the cosmic variance of early massive galaxies. Using its fiducial high-resolution run (their `L1\_m8' run) where the initial mass of the baryonic elements is $m_{\rm gas}\,{=}\,1.34\times10^8\,{\rm M}_\odot$, galaxies of $M_\ast\,{\simeq}\,10^{10}\,{\rm M}_\odot$ at high redshifts are resolved with approximately 100 particles. The model was run on a comoving box of 1\,cGpc$^3$. Given a characteristic volume on the order of $(100\,\mathrm{cMpc})^3$ at $z\,{\gtrsim}\,6$ for large JWST programs like COSMOS-Web (0.54\,deg$^2$; \citealt{Casey2023}) or PRIMER (370\,arcmin$^2$; \citealt{Dunlop2021}), our 1\,cGpc$^3$ simulation contains ${\simeq}\,10^3$ such volumes, allowing us to probe the tail of the cosmic variance distribution out to ${\simeq}\,3\,\sigma$. 

We briefly summarize the key features of the simulation here, while the reader is referred to the original papers for more details. The simulation adopted a hydrodynamical solver \textsc{Swift} \citep{Schaller2024}. Radiative cooling and heating, and multi-phase ISM were modelled based on \citet{PloeckingerSchaye2020} and \citet{SchayeDallaVecchia2008}. Star particles are formed by converting gas particles, to follow the Kennicutt-Schmidt law \citep{SchayeDallaVecchia2008}. Stellar feedback including supernova (SN) and winds was implemented using the models of \citet{Chaikin2023}. The BHs are seeded with the intial subgrid mass of $10^5\,{\rm M}_\odot$ in haloes with the total mass greater than $2.8\times10^{11}\,{\rm M}_\odot \times (m_{\rm gas}/1.07\times10^9\,{\rm M}_\odot)$. As in \citet{Springel2005}, the BHs grow at a modified Bondi-Hoyle accretion rate, via mass transfer from the neighboring gas particles, and through mergers \citep{Bahe2022}. To compensate for the numerical resolution, however, their growth is enhanced in dense regions following \citet{BoothSchaye2009}. The BHs are repositioned at each time step to the local potential minima, since otherwise the lack of dynamical friction due to the resolution fails to hold the BHs to the galaxy center, substantially reducing the impact of AGN feedback \citep{Bahe2022}. The fiducial model used for our analysis implemented a `thermal-mode' AGN feedback of \citep{BoothSchaye2009}, while a `kinetic-mode' jet-like feedback model is also available for lower-resolution simulations \citep{Husko2022}. The subgrid model parameters were calibrated to the observed SMF and cluster gas fractions at $z\,{\simeq}\,0$, by a machine learning approach using Gaussian process emulators, as detailed in \citet{Kugel2023}. The strength of the AGN feedback is determined by a single parameter, $\Delta T_{\rm AGN}$, the minimum anticipated increase in the temperature of the gas particle set for the feedback energy to be released. $\Delta T_{\rm AGN}\,{=}\,10^{8.07}\,{\rm K}$ for the fiducial model. Finally, the simulation assumes a \citet{Chabrier2003} IMF and the Dark Energy Survey year 3 \citep[DES Y3;][]{Abbott2022} cosmology. 

\subsection{The simulated galaxies and properties}
\label{ssec_catalog}

We use the catalogs of the simulated galaxies and haloes of FLAMINGO, constructed by applying \textsc{VELOCIraptor} finder \citep[VR;][]{Elahi2019} to the snapshots. The haloes were identified via a FoF algorithm with a linking length of 0.2 times that of the mean particle separation. Subhaloes are identified via FoF on a six-dimensional phase space using both the positional and kinematic information. Central galaxies are identified as the most distinct structure on the phase space, while the rest are defined as satellites. The halo and galaxy properties are computed using the Spherical Overdensity and Aperture Processor (SOAP; \citealt{McGibbon2025}). While the SOAP catalogs provide the galaxy properties processed within various apertures, we adopt those integrated within the 3-D radius of 50\,pkpc from the center, as it is the aperture with which the model was calibrated. Our fiducial definition of the halo mass (and similarly for other halo properties) for later analysis is $M_{\rm h}\,{=}\,M_{200}$, i.e. the total mass encompassed within the radius ($R_{200}$) within which the mean density is 200 times the critical density. When halo properties should be considered together with associated galaxy properties, such as when computing the stellar and baryon fraction, we link the properties of the central galaxy that a halo hosts to the halo property, and ignore satellites. 

We also make use of the simulation merger tree, to track the evolution of the massive galaxies from redshift of about 10 to 7. The merger tree, which is part of the FLAMINGO data products, was constructed following \citet{Jiang2014}. It identifies progenitors and descendants by tracing the most bound particles across snapshots. 

\subsection{Construction of subboxes}
\label{ssec_subbox}

We divide the entire (1\,cGpc)$^3$ FLAMINGO volume into subboxes of a given size. Our fiducial subbox volume is $(100\,\mathrm{cMpc})^3$ (${=}\,10^6\,\mathrm{cMpc}^3$), chosen to be representative of larger-area JWST surveys such as COSMOS-Web \citep{Casey2023} and PRIMER \citep{Dunlop2021}, as noted earlier. This provides a conservative baseline, as the impact of CV may be even more pronounced in the smaller volumes of deep pencil-beam surveys like JADES. However, we also checked and present the results for two more volumes, $3\times10^5\,{\rm cMpc}^3$ and $10^7\,{\rm cMpc}^3$, where the prediction is sensitive to the volume. For each choice of volume, we divide the simulation box into cubes of the given size, along its 3-D positional axes without any overlap between them. This results in a total of 1,000 unique subboxes of (100\,cMpc)$^3$ in volume for the fiducial choice. 

\begin{figure*}
\includegraphics[width=1.0\linewidth]{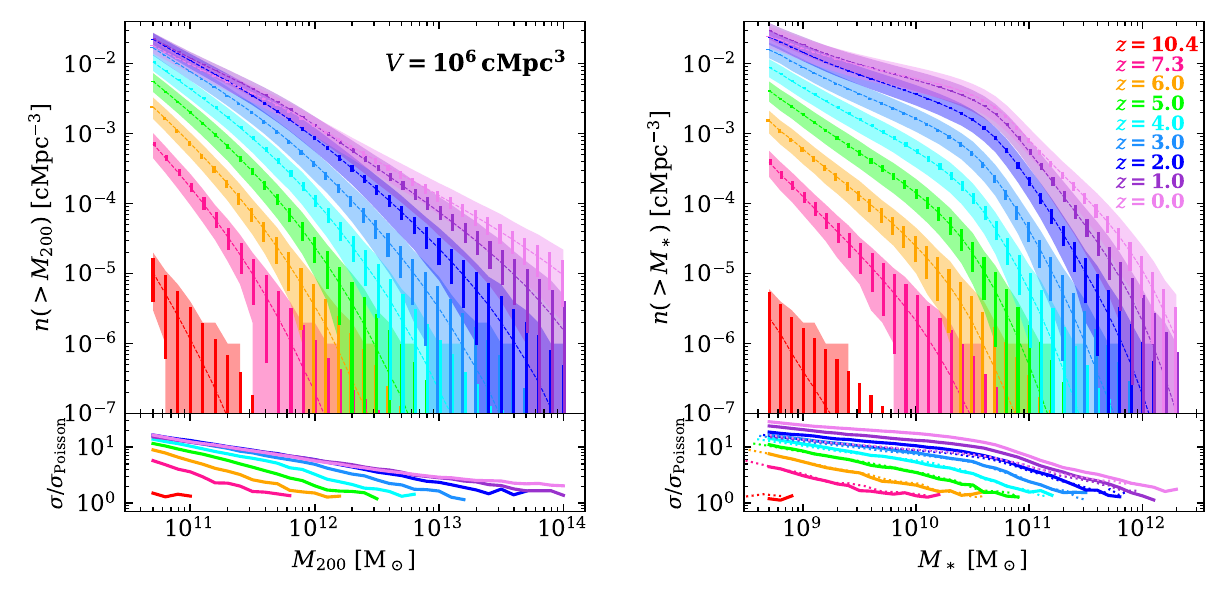}
\vspace{-0.5cm}
\caption{Cumulative halo (left) and stellar (right) mass functions from the FLAMINGO simulation (L1\_m8) over $0\,{<}\,z\,{<}\,10$ (dashed line), together with the 95th-percentile scatter (shaded bands) measured across 1,000 independent $(100\,\mathrm{cMpc})^3$ sub-volumes. For comparison, the error bars show the corresponding variance from a Poissonised reference sample, in which galaxy positions were randomised to remove the intrinsic clustering. The lower panels plot the ratio of the dispersions between the fiducial and Poisson cases (only for the bins with more than three objects per subbox). The results highlight that cosmic variance contributes a substantial excess over Poisson noise, with the relative impact increasing towards lower masses and lower redshifts. Also, the dotted lines in the lower right panel show the ratio for halo mass, matched to stellar mass via the mean SMHMR. The cosmic variance is greater for the stellar mass function than for the halo mass function at $M_\ast\,{\lesssim},10^{11}\,{\rm M}_\odot$ and $z\,{\lesssim}\,2$, an effect we attribute to the inclusion of satellites (see text).}
\label{fig_MFs}
\end{figure*}

\subsection{Definition of samples}
\label{ssec_sample}

Throughout our analysis, we consider all galaxies identified in the SOAP catalog, which contains all central (satellite) galaxies with 32 (20) or more particles including all particle types. Most of our analysis focuses on massive galaxies (namely, $M_\ast\,{\simeq}\,10^{10}\,{\rm M}_\odot$ galaxies at $z\,{\simeq}\,7$), which are typically resolved with more than a hundred baryonic particles. Particularly, we center our investigation on $M_{\rm h,max}$ and $M_{\rm \ast, max}$ galaxies, which we define as the most massive halo and galaxy in each subbox, respectively. 

To evaluate the CV arising from the large-scale density fluctuations in excess of a Poisson noise, we also construct a Poissonian reference sample, where the positional information of the simulated galaxies are redrawn randomly from a uniform distribution, while other properties remain unaltered. 

Stellar mass estimates commonly adopt a conservative uncertainty of at least 0.3 dex, reflecting the characteristic floor set by systematic errors in stellar population models (e.g., \citealt{Conroy2009}). However, we \textit{do not} apply an arbitrary scatter to the mass of the simulated galaxies to account for that, as our goal is to evaluate the intrinsic strength and impact of the CV on the galaxy evolution. We find that the masses such as $M_{\rm h,max}$ and $M_{\rm \ast, max}$ are increased approximately by the same factor when the observational uncertainties are taken into account by introducing a scatter.

\section[result]{Results}
\label{sec_result}

The cosmic variance, and the large-scale coherence in the early galaxy evolution (if it exists), can be revealed in several measurements, including the mass functions, stellar mass fraction ($f_\ast\,{=}\,M_\ast\,/\,(M_{\rm 200}f_{\rm b,cosmic})$), and SMHMR. In this section, we present the FLAMINGO predictions on the halo mass function (HMF) and SMF (Sect.~\ref{ssec_MFs}), the properties of the most massive objects in each survey volume (Sect.~\ref{ssec_Mmaxs}), and the SMHMR and baryon fraction in environments using various indicators (Sect.~\ref{ssec_SMHM} and \ref{ssec_fb}, respectively), as well as their redshift evolution (Sect.~\ref{ssec_evol}). 

\begin{figure*}
\includegraphics[width=1.0\linewidth]{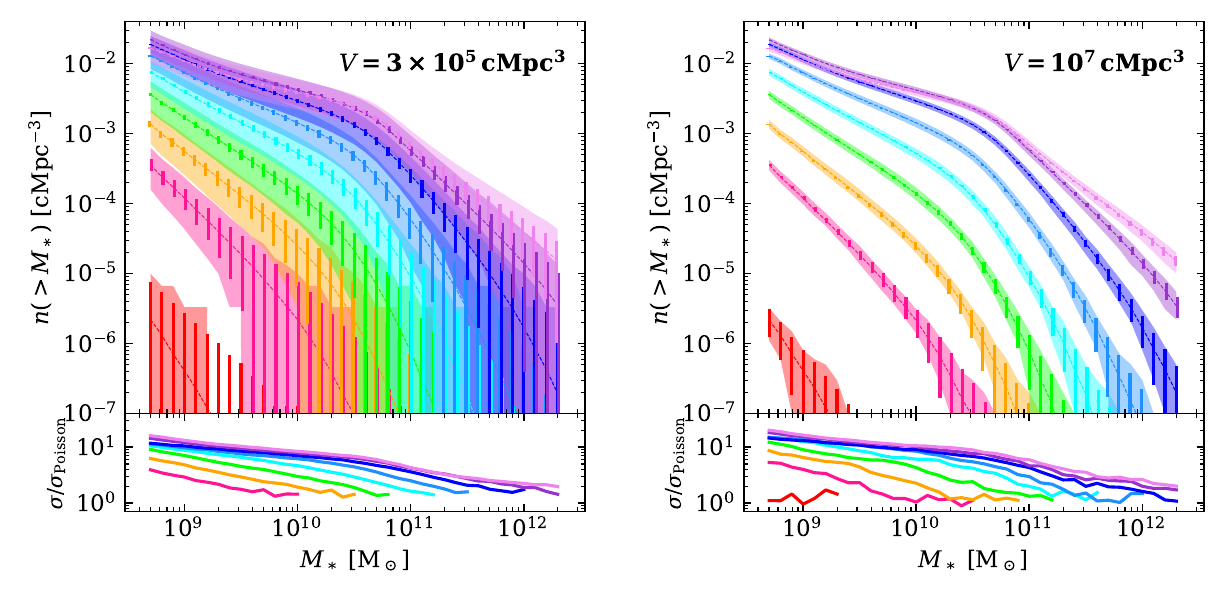}
\vspace{-0.5cm}
\caption{Stellar mass functions, as in Fig.~\ref{fig_MFs}, but measured in sub-volumes of $3\times10^5\,\mathrm{cMpc}^3$ (left panel) and $10^7\,\mathrm{cMpc}^3$ (right panel). }
\label{fig_MFs_vol}
\end{figure*}

\subsection{The cosmic variance of the mass functions}
\label{ssec_MFs}

One of the most basic properties subject to CV is the HMF. Given the initial conditions and the assumed cosmology, $N$-body simulations return accurate predictions for the HMF, as the formation of haloes and growth of large-scale structures are almost solely influenced by gravity. Such predictions are only subject to the uncertainties in the halo definition, numerical resolution, and the variance in the largest-scale mode as restricted by the box and represented in the initial conditions. The left panel of Fig.~\ref{fig_MFs} presents the cumulative HMF from $z\,{\simeq}\,10$ to 0, predicted by the fiducial FLAMINGO simulation, together with the 2\,$\sigma$ variance among 1,000 subboxes of (100\,cMpc)$^3$ as indicated by the bands. 

We follow \citet{Moster2011} and \citetalias{Jespersen2025a} in defining cosmic variance as the additional variance in galaxy number counts beyond those from a Poisson distribution, due to the large-scale mode of clustering. A Poisson distribution is expected when there is no clustering of objects in the parent distribution. Studies have demonstrated variance due to the large-scale mode from both simulations and observations in excess of Poisson \citep[e.g.,][]{Steinhardt2021}. The distribution of the variance has been shown to be well approximated by a gamma function \citep[e.g.,][]{Steinhardt2021, Jespersen2025a}. 

As noted in Sect.~\ref{ssec_sample}, we also construct `Poissonian' samples, to evaluate the variance in excess of a Poissonian expectation where the large-scale clustering is only due to chance, not caused by genuine underlying matter fluctuations. The HMF from the Poissonian samples are also presented in Fig.~\ref{fig_MFs} for comparison, with the 2\,$\sigma$ ranges. There are two trends revealed: the variance in excess of the Poissonian is greater at 1) lower mass, and 2) lower redshifts, as can be clearly seen in the lower panel that shows the ratio of the dispersion between the fiducial and the Poisson samples. The greater relative importance of CV at lower mass is consistent with the finding of \citetalias{Jespersen2025a}, who demonstrated that at the massive end, $\sigma_{\rm CV}N\,{<}\,\sqrt{N}$. In this expression, following \citet{Moster2011} and \citetalias{Jespersen2025a}, $\sigma_{\rm CV}$ is the fractional standard deviation, so $\sigma_{\rm CV} N$ gives the absolute standard deviation in the number count due to CV. This is compared directly to $\sqrt{N}$, the absolute standard deviation from Poisson sampling. The inequality $\sigma_{\rm CV} N\,{<}\,\sqrt{N}$ therefore means the CV is sub-dominant, and the total variance is in the `Poisson regime'. In contrast, at lower mass, where $N$ is greater, $\sigma_{\rm CV}N\,{>}\,\sqrt{N}$, indicating that CV dominates. The second trend of the greater CV at lower redshifts, simply reflects the growth of the large-scale mode fluctuations with time due to the evolution of the matter field in the non-linear regime. The combination of these findings indicates that the variance of the halo distribution \textit{cannot} serve as a solution to the tension of the early massive galaxy abundance with LCDM, whose variance is rather dominated by Poisson noise. Nevertheless, the total variance including CV at $z\,{\simeq}\,7$ in the number of count of $M_{200}\,{\simeq}\,10^{11}\,{\rm M}_\odot$ haloes is 3 to 4 times greater than the expectation in case of no underlying clustering. 

Being affected by baryonic processes, the SMFs can provide additional information about the coherence in the galaxy evolution on large scales, complementary to the HMFs. The cumulative number counts at a given stellar mass, alongside their 2\,$\sigma$ variance, are presented in the right panel of Fig.~\ref{fig_MFs}. While overall the two trends previously found for the HMFs, i.e. the greater relative importance of CV for the lower-mass, lower-redshift galaxies, are also found for the SMFs, there is increased CV (by up to a factor of 2) compared to the HMFs at same redshifts and mass (matched via the SMHMR; dotted lines), particularly at $z\,{\lesssim}\,2$. This could be due to the fact that the HMFs only take into account isolated haloes, while satellite galaxies (associated with subhaloes) are also accounted for in the SMFs. We verified that there is no indication of an increased CV for the SMF when only centrals are included. The inclusion of satellites thus increases the number of samples, pushing the statistics into the CV-dominated regime ($\sigma_{\rm CV}N\,{>}\,\sqrt{N}$). This explains why the increase in the CV of SMF relative to that of HMF is greater at lower mass and lower redshift, where satellites dominate the galaxy count. Otherwise, the differences between the HMFs and SMFs are so marginal that it is hard to probe whether there is additional (or suppressed) variance in the SMFs relative to the HMFs because of any large-scale coherence in the baryonic processes. 

Studies have found that the total variance in the mass functions including CV follows a gamma function \citep[e.g.,][]{Steinhardt2021, Jespersen2025a}. We revisit this by fitting the full distribution of the HMFs and SMFs with gamma functions across mass and redshift. We confirm that the overall fit is acceptable, while that with a Poissonian is not (Appendix~\ref{sec_appA}). We provide the best-fitting parameters in Table~\ref{tab_nM200} and \ref{tab_nMs} of Appendix~\ref{sec_appA}. 

\begin{figure*}
\includegraphics[width=1.0\linewidth]{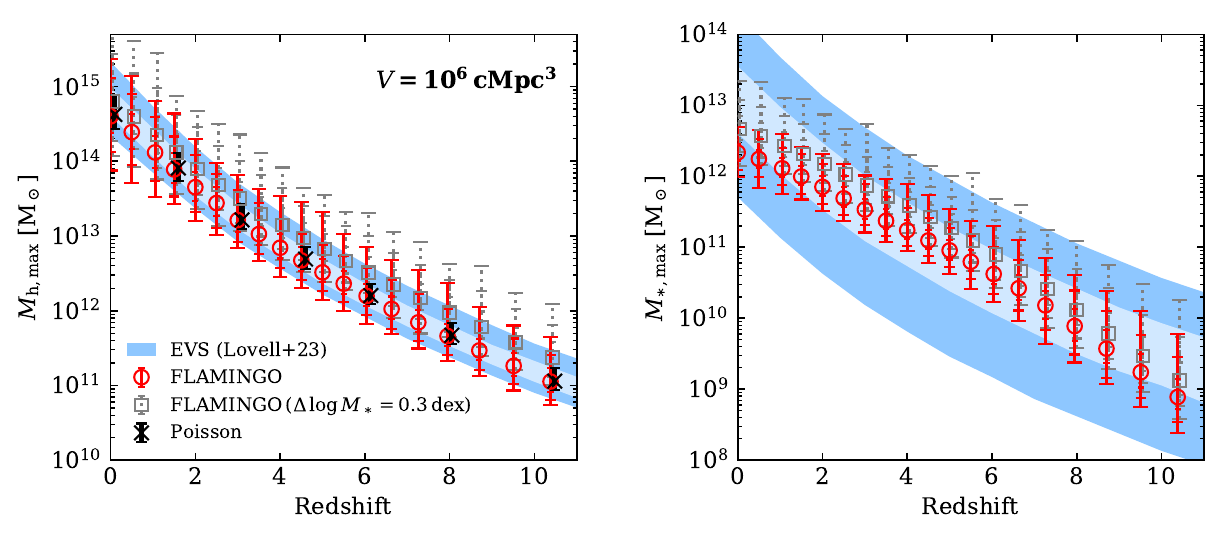}
\vspace{-0.5cm}
\caption{Cosmic variance of the most massive halo ($M_{\rm h,\max}$; left) and galaxy ($M_{\ast,\max}$; right) in $(100\,\mathrm{cMpc})^3$ sub-volumes of the FLAMINGO simulation (L1\_m8) over $0\,{<}\,z\,{<}\,10$. Circles mark the medians, and vertical error bars show the 68th, 95th, and 99.7th percentiles across 1,000 sub-volumes. Grey squares with error bars indicate the same quantities after adding a Gaussian random scatter of 0.3\,dex to mimic observational mass uncertainties (shifted by $\Delta z\,{=}\,0.05$ for clarity). Predictions from Extreme Value Statistics (EVS; \citealt{Lovell2023}) are shown for comparison, with shaded bands denoting the 68th and 99.7th percentile ranges. EVS underestimates the variance in $M_{\rm h,\max}$ because it accounts only for Poisson noise and neglects cosmic variance. This is directly confirmed by the results from the reference Poissonian samples (black crosses with error bars representing the 68th percentile range), which show excellent agreement with the EVS result. The broader dispersion in $M_{\ast,\max}$ for EVS arises from the wider stellar-fraction distribution assumed by \citet{Lovell2023} relative to that predicted by FLAMINGO (see Fig.~\ref{fig_fs}). }
\label{fig_Mmax}
\end{figure*}

We also compare the SMF and their variance for survey volumes covering $3\times10^5\,{\rm cMpc}^3$ to $10^7\,{\rm cMpc}^3$ in Fig.~\ref{fig_MFs_vol}. The variance in the matter distribution decays when averaged over larger scales, resulting in the hierarchical structure formation \citep[e.g.,][]{BBKS, LaceyCole1993}. Therefore, a naive expectation is for the CV to decrease with the increasing volume of the subboxes. However, the results show no strong indication that the CV (in addition to the Poisson variance) decreases with the volumes considered here. Rather, the CV increases monotonically with volume at $z\,{\gtrsim}\,2$. This is because the increase in the number of galaxies in the larger volume pulls the SMFs away from the Poisson-dominated regime (where $\sigma_{\rm CV}N\,{<}\,\sqrt{N}$), compensating for the reduced box-scale density fluctuations. This indicates the advantage of combining multiple independent fields over a single wider-field survey of the same total area.

\subsection{Properties of the most massive galaxies in JWST-like survey volumes}
\label{ssec_Mmaxs}

The discovery of the massive galaxies or structures such as galaxy overdensities and protoclusters at early times of $z\,{\gtrsim}\,5$, mainly driven by JWST \citep[e.g.,][]{Labbe2023, Helton2024b}, has been often claimed to be in tension with LCDM or galaxy formation model \citep[e.g.,][]{Boylan-Kolchin2023, Lovell2023}\footnote{Subsequent investigations have revealed that many of these massive galaxy candidates may have significant AGN contributions, casting doubt on their stellar mass estimates \citep[e.g.,][]{Fujimoto2024, WangB2024, D'Eugenio2025}. Furthermore, the aperture effect in identifying protocluster candidates significantly impacts the assessment of their mass and the resulting apparent tension \citep{Lim2024}.}. However, robustly evaluating such claims requires a precise accounting of CV, which remains a significant challenge at high redshifts. Particularly, what information can be extracted about the galaxy formation and cosmology from the observations of only one or a few brightest objects in a relatively small survey volume is not understood clearly. Furthermore, even the stellar mass fraction and SMHMR derived for the average population at lower redshifts are often times assumed for the early massive galaxies to infer their properties in the literature. In this section, we validate these assumptions and explore the properties of the most massive galaxies with the FLAMINGO predictions. 

\begin{figure*}
\includegraphics[width=1.01\linewidth]{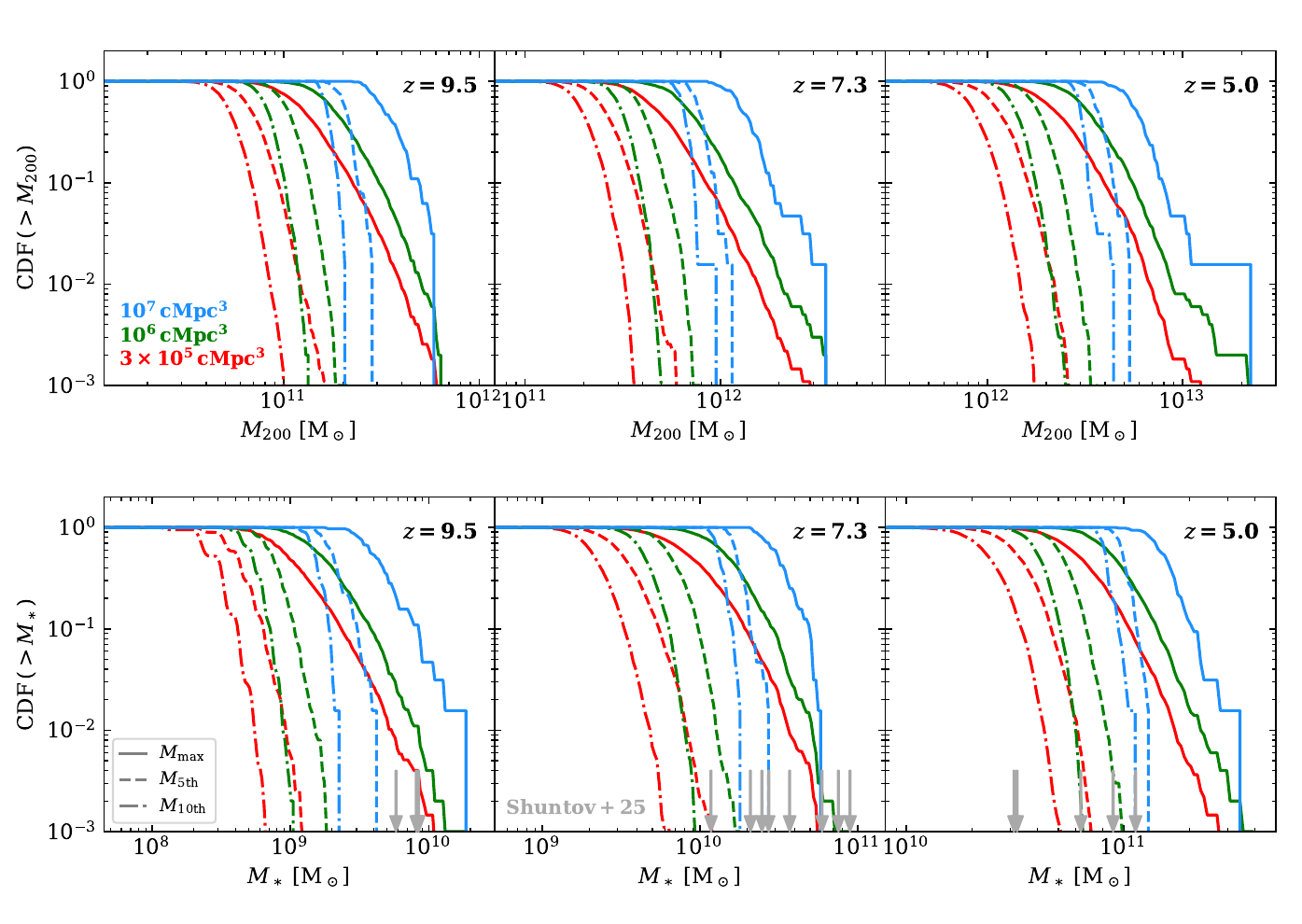}
\vspace{-0.7cm}
\caption{Cumulative distributions of halo mass ($M_{200}$; top row) and stellar mass ($M_{\ast}$; bottom row) for the most massive (solid lines), 5th most massive (dashed lines), and 10th most massive (dot–dashed lines) objects in sub-volumes of $3\times10^5\,\mathrm{cMpc}^3$ (red), $10^6\,\mathrm{cMpc}^3$ (green), and $10^7\,\mathrm{cMpc}^3$ (blue) from the FLAMINGO simulation (L1\_m8) at $5\,{<}\,z\,{<}\,10$. When comparing to observations, random scatter should be convolved to mimic mass-estimate uncertainties. For reference, we include the most massive high-$z$ galaxies observed in COSMOS-Web from \citet{Shuntov2025}. The survey volume roughly corresponds to $4\times10^6\,\mathrm{cMpc}^3$ for $z\,{=}\,5$--9. We only present their samples with JWST MIRI photometry, which results in more robust stellar mass estimates. They also corrected for the Eddington bias in their stellar mass estimates. }
\label{fig_Mth}
\end{figure*}

\subsubsection{$M_{\rm h, max}$ and $M_{\rm \ast, max}$}

First, we explore the halo and stellar masses of the most massive haloes and galaxies in the given survey volumes, $M_{\rm h, max}$ and $M_{\rm \ast, max}$, respectively. The results are shown in Fig.~\ref{fig_Mmax} with the median, 1\,$\sigma$ (the 68th percentile range), 2\,$\sigma$ (the 95th percentile), and 3\,$\sigma$ (the 99.7th percentile) ranges from 1,000 subboxes of (100\,cMpc)$^3$. The mass of the most massive objects can vary from field to field typically by an order of magnitude. This demonstrates a potentially huge impact of the CV on those observations, which is usually omitted or underestimated. Notably, the variance is greater for $M_{\rm \ast, max}$ than for $M_{\rm h, max}$ at $z\,{\gtrsim}\,6$. This reflects an additional large-scale coherence in the galaxy formation and baryonic process, apart from the large-scale density fluctuation, as demonstrated in Sect.~\ref{ssec_SMHM}. Additionally, we also present the results where a Gaussian random scatter of 0.3\,dex was introduced to account for an observational uncertainty in mass estimates. The effect is an overall shift of the $M_{\rm max}$ distribution to a higher mass by a similar amount at all redshifts. 

We also compare the prediction with that from the Extreme Value Statistics \citep[EVS;][]{Lovell2023} in Fig.~\ref{fig_Mmax}. The EVS under-predicts the variance in $M_{\rm h, max}$ over all redshift range, particularly at high redshifts of $z\,{\gtrsim}\,3$. Since the EVS only takes into account Poisson variance but not CV, the discrepancy here directly demonstrates the impact of the large-scale density fluctuations on $M_{\rm h, max}$. We also confirm that the results from our Poissonian samples are consistent with the EVS, under-estimating the variance of $M_{\rm h, max}$, which rules out that the difference could be due to other uncertainties such as the assumed HMFs and cosmology. In the right panel, the EVS predicts a much broader distribution for $M_{\rm \ast, max}$. This breadth, however, stems from the large dispersion in stellar fraction assumed by \citet{Lovell2023}, not from the EVS framework itself. FLAMINGO also predicts a stellar fraction for the most massive galaxies that evolves strongly with redshift, resulting in lower values at $z\,{\gtrsim}\,8$ and $z\,{\lesssim}\,3$ (as will be discussed later). Despite this difference in the underlying model for the stellar fraction, the EVS and FLAMINGO predictions are in good agreement within the 1\,$\sigma$ variance. 

In Fig.~\ref{fig_Mth}, we predict the masses of the top 10 most massive galaxies expected within survey volumes of $3\times10^5$ to $10^7\,\mathrm{cMpc}^3$ at $z\,{\simeq}\,5, 7.3,$ and $9.5$, facilitating a direct comparison with observations of rare systems. We compare our predictions with the massive galaxy candidates from \citet{Shuntov2025}, whose JWST MIRI photometry provides robust stellar masses by mitigating the 0.2--0.6\,dex overestimation common in its absence \citep{Papovich2023, Williams2024, WangT2025}. The survey volumes of \citet{Shuntov2025} (${\simeq}\,1\times10^{6}\,\mathrm{cMpc}^3$ for each redshift shown here) are comparable to our fiducial subboxes.

At $z\,{\simeq}\,5$, the observations are consistent with FLAMINGO, with the most massive objects existing with ${\gtrsim}\,30\%$ probability. However, a significant tension emerges at $z\,{\gtrsim}\,7$, where the observed galaxies are more massive than predicted, exhibiting existence probabilities of ${\lesssim}\,0.1\%$. This tension is subject to uncertainties, particularly in photometric redshifts and stellar mass estimates; although \citet{Shuntov2025} correct for Eddington bias, systematic uncertainties in stellar population modeling may remain. A robust comparison requires convolving our predictions with typical observational mass uncertainties (${\simeq}\,0.3$\,dex), as the steep high-mass slope of the mass function makes probabilities highly sensitive to such errors.

\begin{figure*}
\includegraphics[width=0.92\linewidth]{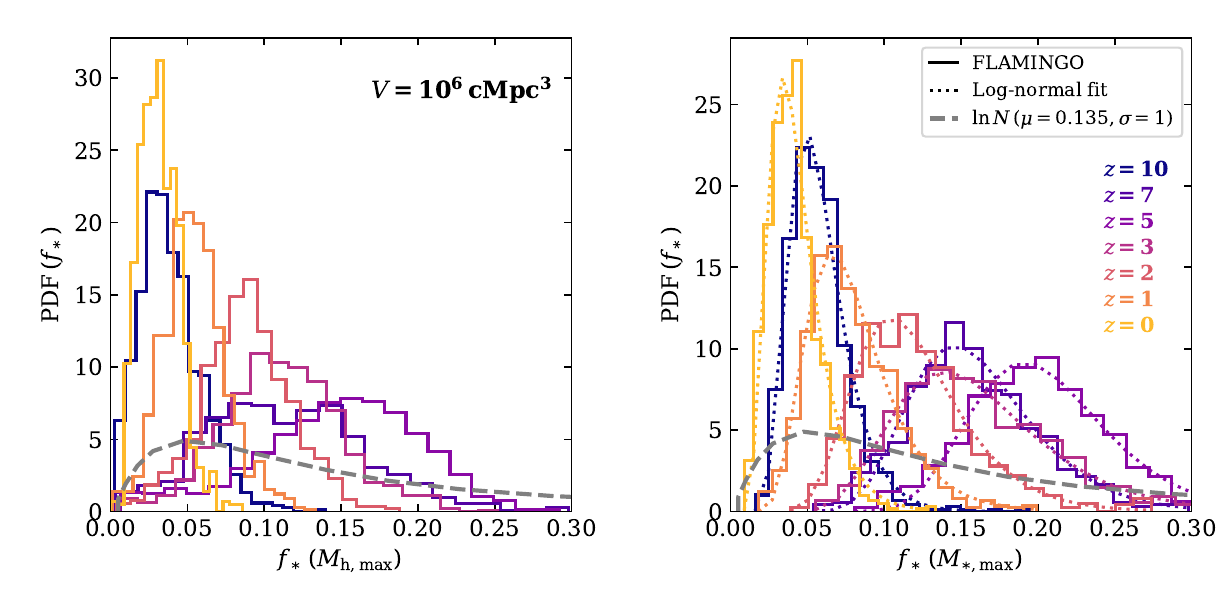}
\vspace{-0.2cm}
\caption{Distribution of the stellar fraction, $f_\ast\,{=}\,M_\ast\,/\,(M_{200}f_{\rm b,cosmic})$, for the most massive halo (left) and galaxy (right) in each of 1,000 $(100\,\mathrm{cMpc})^3$ sub-volumes of the FLAMINGO simulation (L1\_m8). Dotted curves show log-normal fits to the simulated distributions (the best-fitting parameters are summarized in Table~\ref{tab_fs}). For comparison, the grey dashed line indicates a commonly adopted log-normal model with $\mu\,{=}\,0.135$ and $\sigma\,{=}\,1$ \citep[e.g.,][]{Lovell2023}. Relative to this assumption, the FLAMINGO predictions yield narrower distributions with higher mean stellar fractions.}
\label{fig_fs}
\end{figure*}

\begin{figure*}
\includegraphics[width=0.99\linewidth]{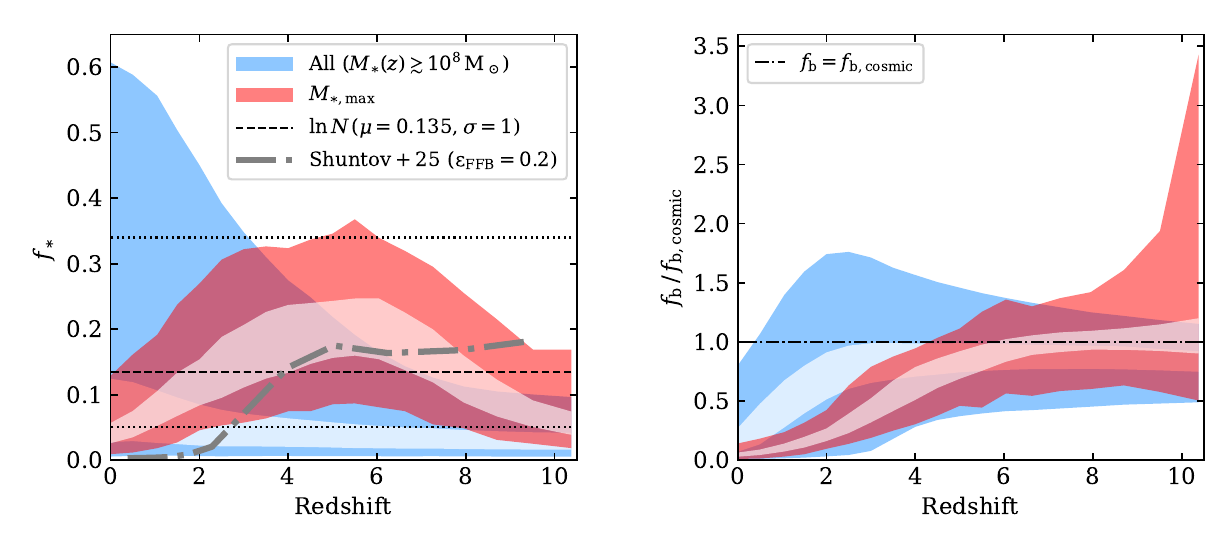}
\vspace{-0.2cm}
\caption{Redshift evolution of the stellar fraction of the most massive galaxy (left panel) and the baryonic fraction of its host halo (right panel) in each of 1,000 $(100\,\mathrm{cMpc})^3$ sub-volumes of the FLAMINGO simulation (L1\_m8), compared to the full galaxy population (resolved down to $M_\ast\,{\simeq}\,10^8\,\mathrm{M}_\odot$). Shaded bands show the 68th and 99.7th percentile ranges. In the left panel, the grey dashed and dotted lines denote the median and 68th percentile range of a commonly assumed log-normal model with $\mu\,{=}\,0.135$ and $\sigma\,{=}\,1$. In the left panel, also shown is the prediction from the feedback-free starburst (FFB) model \citep{Dekel2023,Li2024}, with $\epsilon_{\rm FFB}\,{=}\,0.2$, adopted from \citet{Shuntov2025} and evaluated for a typical halo mass near $M_{\rm h,\max}$ in FLAMINGO. At $z\,{\gtrsim}\,4$, the FLAMINGO predictions are in good agreement with the FFB model.}
\label{fig_fs_fb_z}
\end{figure*}

\subsubsection{Stellar mass fraction}

Many studies of early massive galaxies, and the impact of CV, have assumed a distribution of the stellar mass fraction, $f_\ast\,{=}\,M_\ast\,/\,(M_{200}f_{\rm b,cosmic})$, derived either empirically from observations \citep[e.g.,][]{Finkelstein2015, Harikane2016, Harikane2018, Stefanon2021, Jespersen2025a} or based on simple halo models \citep{Tacchella2018}. $f_{\rm b,cosmic}\,{=}\,0.16$ is the universal baryon fraction, as determined by CMB experiments like Planck. $f_\ast\,{=}\,1$ corresponds to the case where all expected baryon mass is found in stars. 

In Fig.~\ref{fig_fs}, we show the full distribution of $f_\ast$ for $M_{\rm h,max}$ and $M_{\rm \ast, max}$ galaxies from 1,000 subboxes of (100\,cMpc)$^3$. First, it is seen that the distribution begins at $z\,{\simeq}\,10$ with a low average SFE of $f_\ast\,{\simeq}\,0.05$, i.e. only about 5 per cent of the baryons are converted to stars. But the evolution is rapid such that at $z\,{\simeq}\,7$ the distribution becomes broad, centered around $f_\ast\,{\simeq}\,0.1$--0.15. The stellar fraction settles by redshift of about 5, peaking at around 20 per cent on average, before it declines as it approaches $z\,{=}\,0$ (a trend we attribute to late-time feedback processes, as discussed in Sect.~\ref{sec_discussion}). 

Studies such as \citet{Lovell2023} adopted a normal or log-normal distribution for $f_\ast$ of the early-time massive galaxies. While we find, in Fig.~\ref{fig_fs}, that log-normal distributions fit well the FLAMINGO predictions, they are found to be much narrower and peak at about five times higher values than assumed in the literature, in addition to the strong, non-trivial redshift evolution. The distributions from the literature fail to match the FLAMINGO predictions for the most massive galaxies, particularly at both high and low redshifts. Normal distributions do not fit the predictions well, as they present the highly skewed tail towards the higher star-formation efficiency. The parameters from the log-normal fitting are summarized in Table~\ref{tab_fs}. 

\begin{table}
 \renewcommand{\arraystretch}{1.2} 
 \centering
  \begin{minipage}{86mm}
  \caption{Log-normal fit to the stellar fraction distribution of the most massive galaxies, ${\it f}_\ast({\it M}_{\rm \ast,max})\,{=}\,\log N(\mu,\sigma)$, from $z\,{\simeq}\,10$ to 0.}
  \begin{tabular}{cccccccc}
\hline
Redshift & 10 & 7 & 5 & 3 & 2 & 1 & 0 \\ 
\hline
\hline
$\mu$ & 0.055 & 0.16 & 0.20 & 0.15 & 0.11 & 0.075 & 0.040 \\
$\sigma$ & 0.33 & 0.26 & 0.22 & 0.32 & 0.31 & 0.36 & 0.41 \\
\hline
\\
\end{tabular}
\vspace{-2mm}
\label{tab_fs}
\end{minipage}
\vspace{-0.2cm}
\end{table}

In Fig.~\ref{fig_fs_fb_z}, we compare the $f_\ast$ distribution of $M_{\rm \ast,max}$ and all simulated galaxies. As shown, the average galaxies have significantly lower stellar fraction than the massive galaxies, with barely any redshift evolution from $z\,{\simeq}\,10$ to 0 apart from the distribution becoming broader with decreasing redshift. The results will be dominated in number by $M_\ast\,{\simeq}\,10^8\,{\rm M}_\odot$ galaxies, the resolution limit of the simulation, as the SMFs increase in normalisation towards low mass across all redshifts (Fig.~\ref{fig_MFs}). This also means that the results do not track the same objects across redshifts. Nevertheless, lack of any significant evolution is notable in comparison with the massive galaxies. The rise and fall of the most massive galaxies in $f_\ast$ with redshift evolution is in broad agreement, at high redshift, with the feedback-free starburst (FFB) model \citep{Dekel2023, Li2024} with $\epsilon_{\rm FFB}\,{=}\,0.2$, adopted from \citet{Shuntov2025}, assuming an average growth history for a halo mass close to $M_{\rm h, max}$ from the FLAMINGO prediction. 

\begin{figure*}
\includegraphics[width=1.\linewidth]{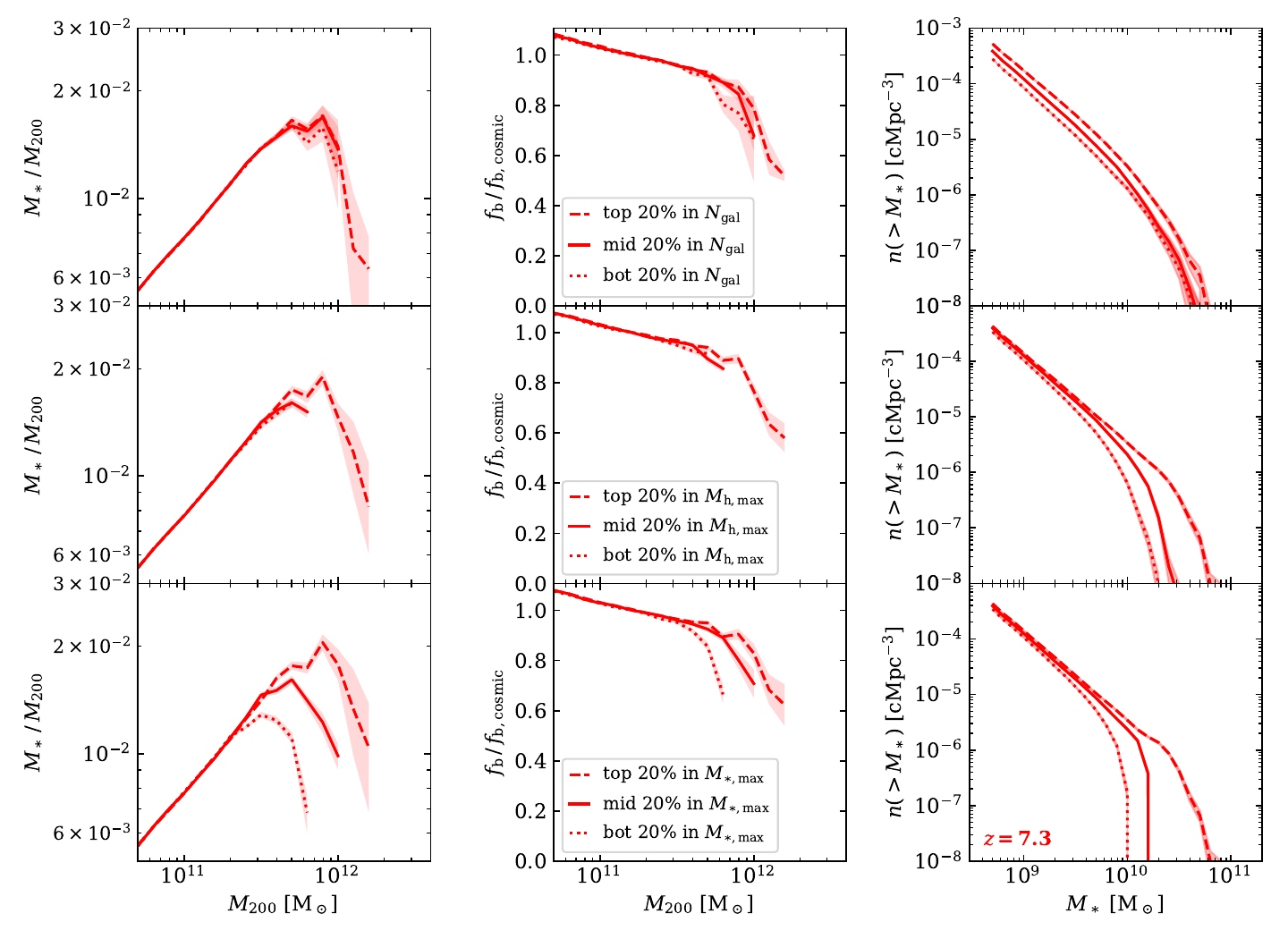}
\vspace{-0.5cm}
\caption{The stellar-to-halo mass relation (SMHMR; left), baryon fraction (middle), and stellar mass function (SMF; right) at $z\,{\simeq}\,7$, measured from 1,000 subboxes of (100\,cMpc)$^3$ in the FLAMINGO simulation (L1\_m8). Results are shown for the top (dashed), middle (solid), and bottom (dotted) 20\% of subboxes, ranked by three different environmental metrics: total galaxy number ($N_{\rm gal}$; top row), maximum halo mass ($M_{\rm h, max}$; middle row), and maximum stellar mass ($M_{\rm \ast, max}$; bottom row); the shaded bands show the 1$\sigma$ error on the mean (10,000 bootstrap samples). The key result is that only the $M_{\rm \ast, max}$ ranking reveals a clear correlation with enhanced galaxy formation efficiency (i.e., a higher SMHMR). In contrast, no correlation is found when ranking by $N_{\rm gal}$, indicating that highly clustered regions do not necessarily host more efficient star formation. The $M_{\rm h, max}$ ranking is inconclusive, as the mid and bottom 20\% subsets probe only the regime $M_{200}\,{\lesssim}\,10^{11.6}\,{\rm M}_\odot$ where the correlation is weak.}
\label{fig_SMHM_z7}
\end{figure*}

\subsubsection{Baryon fraction}

In studies on the early massive galaxies, the baryon fraction, $f_{\rm b}\,{=}\,M_{\rm b}\,/\,M_{200}$, is normally set to $f_{\rm b, cosmic}$, assuming no dependence on environments. Here, $M_{\rm b}$ is the total baryon mass, i.e. the sum of the stellar and gas mass, directly obtained from the simulated (within 50\,pkpc in our analysis; see Sect.~\ref{ssec_catalog}) or observed galaxies, which is not necessarily equal to $M_{200}f_{\rm b,cosmic}$. Also, the baryon fraction at such high redshifts of $z\,{\gtrsim}\,7$ is not well explored. We test these assumptions in Fig.~\ref{fig_fs_fb_z}, in comparison with the average simulated galaxies. As can be seen, the redshift evolution of $f_{\rm b}$ is not straightforward to describe, with multiple phases of increase and decline. The evolution is shaped by a combination of the stellar and AGN feedback as well as the growth of the gravitational potential to prevent the gas from being blown out by feedback. Each process becomes effective at different times, which results in the complicated evolutionary trend. However, the baryon fraction of $M_{\rm \ast, max}$ galaxies remains between 0.8 and 1 times the universal value at $z\,{\gtrsim}\,5$, after which it rapidly decreases towards 0 as approaching $z\,{=}\,0$.

The full distribution of all simulated galaxies, on the contrary, presents a much broader dispersion in general, and a significantly different evolutionary trend at $z\,{\lesssim}\,4$. For both $M_{\rm \ast, max}$ and all galaxies, a fraction of them is shown to have more baryons than the universal value. Most of those are systems that were once satellites during which they lost dark matter from dynamical processes \citep{Lim2025}.

\subsection{The SMHMR in different environments}
\label{ssec_SMHM}

\begin{figure}
\includegraphics[width=1.\linewidth]{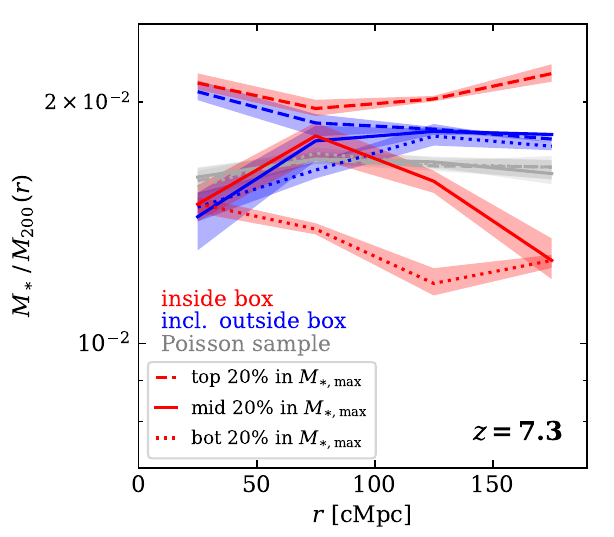}
\vspace{-0.4cm}
\caption{Stellar-to-halo mass relation (SMHMR) at $z\,{\simeq}\,7$ as a function of 3D distance from the most massive galaxy ($M_{\ast,\max}$) in each sub-volume. Curves show the mean SMHMR for the top (dashed), middle (solid), and bottom (dotted) 20\% of 1,000 $(100\,\mathrm{cMpc})^3$ subboxes, ranked by their $M_{\ast,\max}$. We focus on galaxies with $M_{200}\,{=}\,10^{11.5}$--$10^{12}\,\mathrm{M_\odot}$, where environmental differences are strongest (see Fig.~\ref{fig_SMHM_z7}). Red curves include only galaxies that lie both within the radial shell \emph{and} inside the same subbox as the central $M_{\ast,\max}$ galaxy. Blue curves include \emph{all} galaxies in the radial shell, irrespective of subbox. Grey lines show the corresponding results for Poissonian samples with randomized positions. In all cases, the central $M_{\ast,\max}$ galaxy itself is excluded. Shaded regions indicate the $1\sigma$ error on the mean from 10,000 bootstrap samples. A clear $M_{\ast,\max}$–SMHMR correlation persists out to the largest distances probed (${\simeq}\,150\,\mathrm{cMpc}$) when limited to galaxies inside the same subbox, revealing a large-scale coherence in galaxy formation efficiency. However, this signal reduces when galaxies from other subboxes are included, and only appears at radii ${\lesssim}\,100\,\mathrm{cMpc}$ where the selections overlap. The absence of any signal in the Poissonian samples confirms that the coherence is physical rather than a chance clustering effect. }
\label{fig_SMHM_R}
\end{figure}

The stellar-to-halo mass relation is formally defined as $M_\ast\,/\,M_{200}$, which is equivalent to the stellar fraction $f_\ast$ when scaled by the universal baryon fraction, $f_\ast\,{=}\,(M_\ast\,/\,M_{200})\,/\,f_{\rm b, cosmic}$. It is an indicator of the star/galaxy formation efficiency, and is thus one of the most important properties predicted by the galaxy evolution model. SMHMR has been derived empirically from a set of observations up to redshift of about 8 \citep[e.g.,][]{Moster2018, Tacchella2018, Behroozi2019}. However, a common practice in studies of early massive galaxies is to apply the same average relation to infer their properties, without considering its possible dependence on environment. In this section, we quantify the environmental dependence of the SMHMR at $z\,{\simeq}\,7$ by examining how it varies across our simulation subboxes when they are sorted according to two key environmental proxies: (1) the number of galaxies, and (2) the halo and stellar mass of the most massive halo ($M_{\rm h, max}$) or galaxy ($M_{\rm \ast, max}$) within the volume. 

First, we investigate the dependence of the SMHMR on the number of galaxies, $N_{\rm gal}$, within each volume. $N_{\rm gal}$ is limited to the simulation resolution, which is $M_\ast\,{\simeq}\,10^8\,{\rm M}_\odot$. The number of galaxies per volume, by definition, contains the clustering information on the scale of the subbox and above, making it the most natural indicator of environment. We rank 1,000 subboxes of (100\,cMpc)$^3$ according to $N_{\rm gal}$. Figure~\ref{fig_SMHM_z7}, in the top-left panel, shows the average SMHMR for the top, middle, and bottom 20\% subboxes in the rank of $N_{\rm gal}$. As can be seen, there is no evidence of correlation between $N_{\rm gal}$ and SMHMR. This indicates that the galaxy formation efficiency is independent of the environment as traced by clustering, neither enhanced nor suppressed. The grouping of 20\% achieves an optimal signal-to-noise ratio, tested and chosen by trial and error. However, we find no significant changes in our conclusions when averaging over e.g. 33\% for each rank group. In the top-right panel of Fig.~\ref{fig_SMHM_z7}, the cumulative SMFs are shown for each rank group. Although the rank is based on the SMF at the lowest-mass end, which is dominated by the least massive galaxies, the average SMFs for the higher rank group remain higher through the most massive end. We confirm these conclusions including no dependence on the clustering, when using the HMF in lieu of the SMF. 

Next, we explore the correlation of the SMHMR with the most massive galaxies within the volume of (100\,cMpc)$^3$. This is related to the earlier question in Sect.~\ref{ssec_Mmaxs} of what information can be inferred from a few extreme objects about the entire survey footprint from which they were identified. The average SMHMR for the top, middle, and bottom 20\% subboxes as ranked in $M_{\rm h, max}$ are shown in the middle row of Fig.~\ref{fig_SMHM_z7}. The most massive halo in each subbox is excluded when averaging. As shown, we find no evidence for a correlation between the SMHMR and $M_{\rm h, max}$. It should be noted, however, that, for the middle and bottom 20\%, the halo mass is only probed up to $M_{200}\,{\simeq}\,10^{11.6}\,{\rm M}_\odot$. 

However, a strong correlation signal is revealed when the SMHMR and $M_{\rm \ast, max}$ are examined. This is demonstrated in the bottom-left panel of Fig.~\ref{fig_SMHM_z7}, where the average SMHMR was calculated for the subboxes ranked in $M_{\rm \ast, max}$ instead of $M_{\rm h, max}$. The most massive galaxy in each subbox is removed from the averaging. The galaxies are formed about three times more efficiently, at $M_{200}\,{\simeq}\,10^{12}\,{\rm M}_\odot$, in the volumes containing the top 20\% in $M_{\rm \ast, max}$ than in the bottom 20\%. Survey volumes identified with exceptionally bright/massive galaxies or structures are certainly highly biased regions in the clustering and number counts of massive galaxies, as shown in the right panel. Comparing the left and right panels of Fig.~\ref{fig_SMHM_z7}, we conclude that enhanced star formation and SMHMR are found, on average, in volumes with high clustering (particularly of massive galaxies), but not all clustered regions experience a boosted star formation. The most massive galaxy contains unique and remarkable information about the galaxy formation efficiency in an entire survey volume, information that the abundance of objects or the mass of the most massive halo do not provide. This can be useful not only for interpreting observational results, but also for constraining the galaxy formation model. 

\begin{figure*}
\includegraphics[width=1.\linewidth]{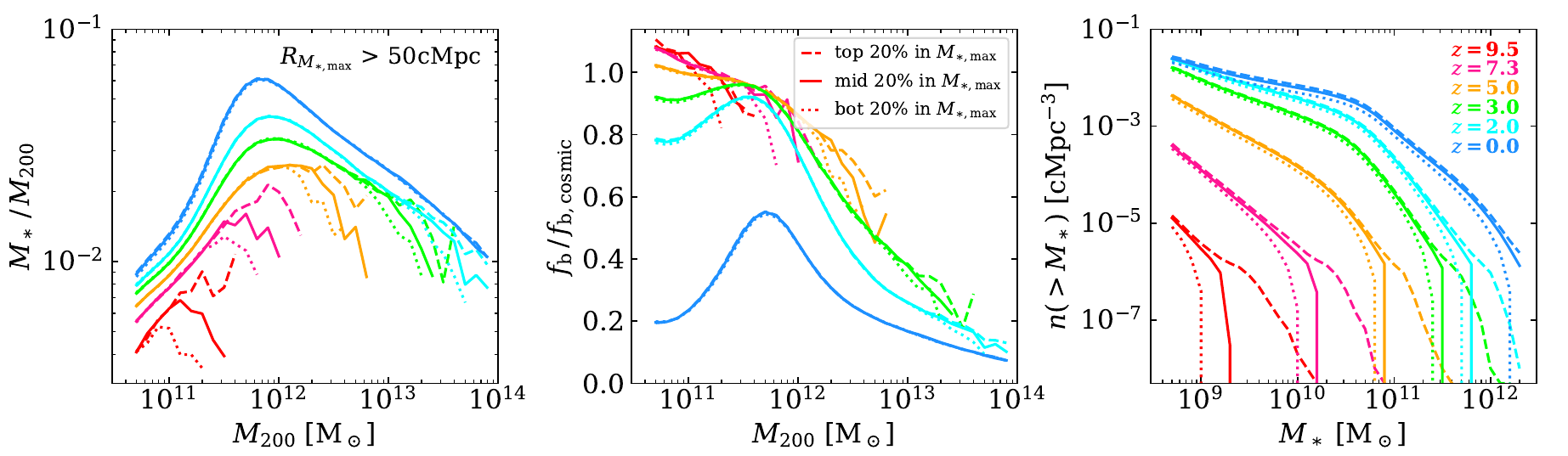}
\vspace{-0.3cm}
\caption{The evolution of the stellar-to-halo mass relation (SMHMR; left), halo baryon fraction (middle), and stellar mass function (SMF; right) from $z\,{\simeq}\,10$ to $z\,{=}\,0$ in the FLAMINGO simulation (L1\_m8), measured from 1,000 subboxes of (100\,cMpc)$^3$. Results are averaged over the top (dashed), middle (solid), and bottom (dotted) 20\% of subboxes ranked by their most massive galaxy's stellar mass ($M_{\rm \ast, max}$). To probe large-scale environmental correlations, only galaxies outside a 3-D aperture of 50\,cMpc from the $M_{\rm \ast, max}$ galaxy are included. The key result is a pronounced redshift evolution in the environmental dependence. The difference in the SMHMR between regions with high and low $M_{\rm \ast, max}$ is strongest at high redshift ($z\,{\gtrsim}\,3$) and weakens significantly towards the present day. }
\label{fig_SMHM_z}
\end{figure*}

To assess the spatial extent of the conformity signal, we measure the SMHMR as a function of three-dimensional distance from the most massive galaxy ($M_{\ast, \max}$) in each $(100\,\mathrm{cMpc})^3$ subbox, out to radii of ${>}\,150\,\mathrm{cMpc}$. We first restrict the measurement to galaxies that lie both within the radial shell and inside the same subbox as the $M_{\ast, \max}$ galaxy. For this selection, shown by the red curves in Fig.~\ref{fig_SMHM_R}, the enhancement in galaxy formation efficiency (i.e. elevated SMHMR) persists out to ${\simeq}\,100\,\mathrm{cMpc}$, demonstrating a striking large-scale galactic conformity. 

However, when \emph{all} galaxies in the radial shell are included---regardless of whether they lie in the same subbox---the signal reduces (blue curves), except at small radii (${\lesssim}\,50\,\mathrm{cMpc}$) where the selections overlap. This behaviour confirms that the detectable coherence is effectively enhanced by the survey footprint rather than extending indefinitely into the surrounding volume. The corresponding Poissonian samples (grey curves), where galaxy positions have been randomized, show no such signal at any radius, ruling out a stochastic clustering origin.

These results imply that the large-scale conformity observed at $z\,{\simeq}\,7$ reflects genuine coherence in baryonic processes within the environmental volume of a massive galaxy, but that this correlation becomes less pronounced once galaxies outside that volume---drawn from distinct large-scale environments---are included. Implications of this footprint-bounded coherence are discussed further in Sect.~\ref{sec_discussion}.

\subsection{Large-scale coherence in baryon fraction}
\label{ssec_fb}

Similarly to SMHMR, we test whether there is a large-scale coherence in the baryon fraction of massive galaxies across (100\,cMpc)$^3$ volume. As shown in the middle column of Fig.~\ref{fig_SMHM_z7}, no evidence of correlation between the SMFs (also the HMFs although not shown) or $M_{\rm h, max}$ and the baryon fraction is present. However, there is a noticeable dependence of $f_{\rm b}$ on $M_{\rm \ast, max}$. The impact of the environment is not as strong as in the case for the SMHMR, with only about 20 per cent less baryon fraction (at $M_{200}\,{\simeq}\,10^{11.7}\,{\rm M}_\odot$) contained in the volumes of the bottom 20\% in $M_{\rm \ast, max}$ compared to the volumes containing the highest $M_{\rm \ast, max}$. Combined with the results in Sect.~\ref{ssec_SMHM}, this implies that the higher (lower) galaxy formation efficiency coherently found within the volumes is primarily due to more (less) star formation per unit baryon mass, as opposed to differences in the total baryon mass fraction. We tested and verified that the conformity in star-formation efficiency is present even at fixed $f_\mathrm{b}$, demonstrating that it is not a secondary, non-linear effect of large-scale variations in baryon abundance.

\subsection{Evolution of the conformity}
\label{ssec_evol}

So far we have only shown the results at redshift of about 7. In Fig.~\ref{fig_SMHM_z}, a full exploration of the SMHMR from $z\,{\simeq}\,10$ to 0 is presented for the top, middle, and bottom 20\% sub-volumes in $M_{\rm \ast, max}$. Only galaxies outside the aperture of 50\,cMpc (but still within the subbox) centered on the most massive galaxy are considered. However, the same conclusions are found with other choices of apertures. For the SMHMR (left), the baryon fraction (middle) and the SMF (right), it is found that the large-scale coherence in galaxy evolution that depends on $M_{\rm \ast, max}$ is strongest at the highest redshift and diminishes at $z\,{\lesssim}\,2$. This indicates that the observations of the ultra-high redshift massive galaxies are expected to be most strongly affected by CV and large-scale conformity in the evolution, requiring careful interpretation to avoid biased conclusions (see also, e.g., \citealt{Jespersen2025b}). This also highlights a continuing need for high-resolution large-volume simulations and accurate modeling of massive galaxies in environments.

\section[discussion]{Discussion}
\label{sec_discussion}

\subsection{The stellar and baryon fraction of the most massive galaxies over cosmic time}

In Sect.~\ref{ssec_Mmaxs}, we demonstrated that the stellar fraction of the most massive galaxy in the survey volume, $f_\ast\,{=}\,M_\ast\,/\,(M_{200}f_{\rm b,cosmic})$, reaches up to about 30\% at $z\,{\simeq}\,2$--$7$---more than five times greater than values in lower-mass galaxies or commonly extrapolated from low-redshift studies (Fig.~\ref{fig_fs}). Furthermore, this fraction exhibits a significant change with redshift, as shown in Fig.~\ref{fig_fs_fb_z}. The baryon fraction of the halo hosting the most massive galaxy, meanwhile, remains near the cosmic value at $z\,{\simeq}\,6$ before declining rapidly at lower redshifts. This trend traces the stellar and baryon fraction of the most massive galaxies identified separately at each redshift, rather than the evolution of individual objects across cosmic time---a distinction supported by studies showing the most massive systems at $z\,{\gtrsim}\,4$ are generally not the progenitors of their $z\,{\simeq}\,0$ counterparts \citep[e.g.,][]{DeLuciaBlaizot2007, Shankar2015, Rennehan2020, Remus2023, Lim2024}. 

The redshift dependence of $f_\ast\,(M_\mathrm{\ast, max})$ is a composite effect, driven by both the growth in the halo mass of these galaxies and an evolution of the relation at fixed halo mass. Specifically, FLAMINGO predicts that at a fixed halo mass, the stellar fraction increases toward lower redshift (Fig.~\ref{fig_SMHM_z}). This fixed-mass evolution produces a stellar fraction that rises rapidly at high redshifts before slowly declining, creating a broad `plateau' in redshift space (Fig.~\ref{fig_fs_fb_z}) where the decline driven by the peak of the SMHMR is counterbalanced by its increasing normalization at fixed mass. This contrasts with simulations like COLIBRE, which predict a largely static fixed-mass relation \citep{Chaikin2025b}, resulting in a SMHMR evolution for the most massive galaxies that is dominated primarily by halo mass growth. 

\begin{figure*}
\includegraphics[width=1.01\linewidth]{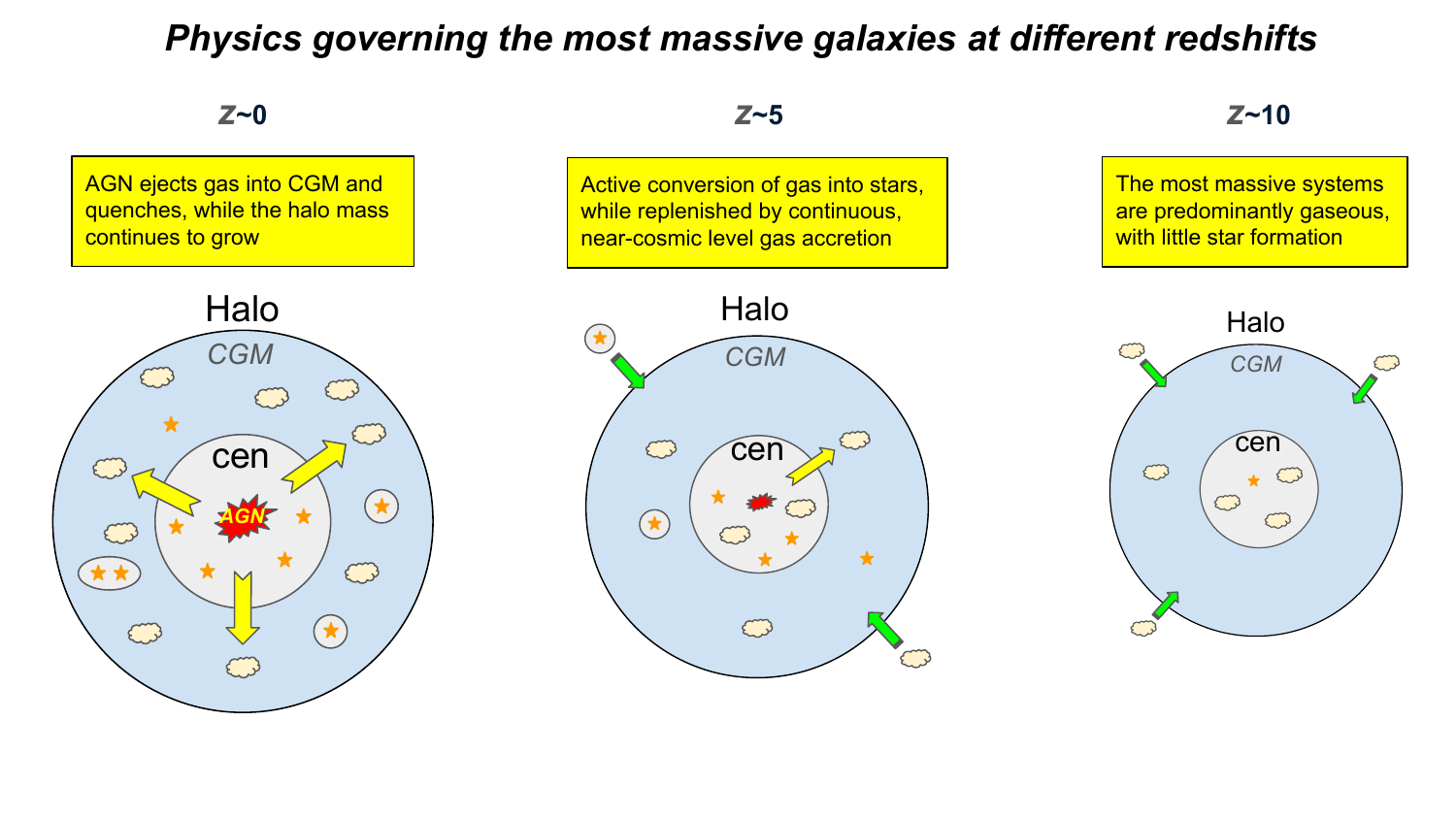}
\vspace{-1.5cm}
\caption{A sketch of the physical state of the most massive galaxies at different redshifts, as predicted by the FLAMINGO simulation (L1\_m8) and inferred from Fig.~\ref{fig_fs_fb_z}. The galaxies shown are identified as the most massive systems individually at each redshift, not as progenitors and descendants. At $z\,{\simeq}\,10$ (right-hand side), the most massive galaxies are gas-dominated systems (white clouds) with inefficient star formation. The most massive systems at $z\,{\simeq}\,5$--$10$ experience efficient star formation (yellow stars), building up their stellar mass while their total halo baryon fraction remains near the cosmic value. The population at $z\,{\simeq}\,5$ is characterized by the onset of powerful AGN feedback (red explosion), which begins to eject gas. Finally, the most massive galaxies at $z\,{<}\,5$ are subject to intensifying AGN activity that quenches them by depleting their gas reservoir, while the growth of the halo mass further reduces the central stellar fraction.}
\label{fig_cartoon}
\end{figure*}

Combining these results with an analysis of the fractions within $R_{200}$, we propose the redshift-dependent picture for the most massive galaxies depicted in Fig.~\ref{fig_cartoon}: by $z\,{\simeq}\,10$, the most massive systems are predominantly gaseous, with little star formation. Subsequently, gas is efficiently converted into stars while being replenished by continuous accretion from the environment. This phase sustains the halo baryon fraction at near-cosmic levels while the stellar fraction rises. At $z\,{\simeq}\,5$, the central black hole in the most massive galaxy becomes sufficiently massive to trigger effective AGN feedback that ejects gas into the circumgalactic medium (CGM). At lower redshifts, these strengthening AGN effects quench the most massive galaxy through gas depletion. Concurrently, the continuous accretion of matter onto the outer halo further reduces the central stellar fraction relative to the total halo mass.

\subsection{Environmental dependence of the SMHMR: physical origin vs. observational effect}

In Sect.~\ref{ssec_SMHM}, we showed that the enhancement in galaxy formation efficiency around the most massive galaxy in a field persists out to the largest distance probed (${\simeq}\,150\,\mathrm{cMpc}$) when the measurement is restricted to galaxies within the same $(100\,\mathrm{cMpc})^3$ subbox. When we instead include \emph{all} galaxies in the spherical shell---i.e.\ including those outside the subbox---the signal is not erased but is significantly weakened for ${\gtrsim}\,50$\,cMpc, as shown by the blue curves in Fig.~\ref{fig_SMHM_R}. Although the original correlated population from the central subbox is still present, it is increasingly outnumbered by galaxies from other subboxes, whose baryonic properties are not correlated with the environment of the target $M_{\ast,\max}$ galaxy. As a result, a reduced conformity signal remains detectable out to ${\lesssim}\,100\,\mathrm{cMpc}$, beyond which it becomes indiscernible. We verified this behaviour using a mass-dependent Voronoi tessellation to define environmental domains and again found that the conformity signal persists, but is weaker, once galaxies beyond the primary volume are included.

These findings suggest that each massive galaxy ($M_{\ast,\max}$) resides at the centre of a coherence volume: a region in which galaxies share both a common Lagrangian footprint and correlated baryonic assembly histories. Identifying a galaxy as the most massive within a survey volume is therefore likely to select its dominant coherence region. Galaxies inside that region remain coherent with the central system and with each other, whereas galaxies beyond it reside within the coherence volumes of other massive systems, and thus weaken the detectable signal when mixed together. Crucially, this framework explains why the observed conformity is footprint-dependent and why including galaxies from uncorrelated environments dilutes---but does not erase---the signal.

To test whether the observed coherence could arise via stochastic clustering rather than a physical correlation, we analysed Poissonian samples with randomized galaxy positions (Sect.~\ref{ssec_sample}). In these data, no conformity signal is detected at any scale, ruling out a purely stochastic origin. This confirms that the large-scale conformity is a genuine physical phenomenon, driven by mechanisms that imprint correlated galaxy formation efficiency across tens of cMpc.

\subsection{Characteristics of the volumes with the highest $M_{\rm \ast, max}$}

As shown in the earlier sections, $M_{\rm \ast, max}$ appears to be a strong indicator of the characteristics of a given survey volume. However, another interesting question is what may have led to the formation of such high-$M_{\rm \ast, max}$ galaxies in those boxes in the first place. 

\begin{figure*}
\includegraphics[width=1.\linewidth]{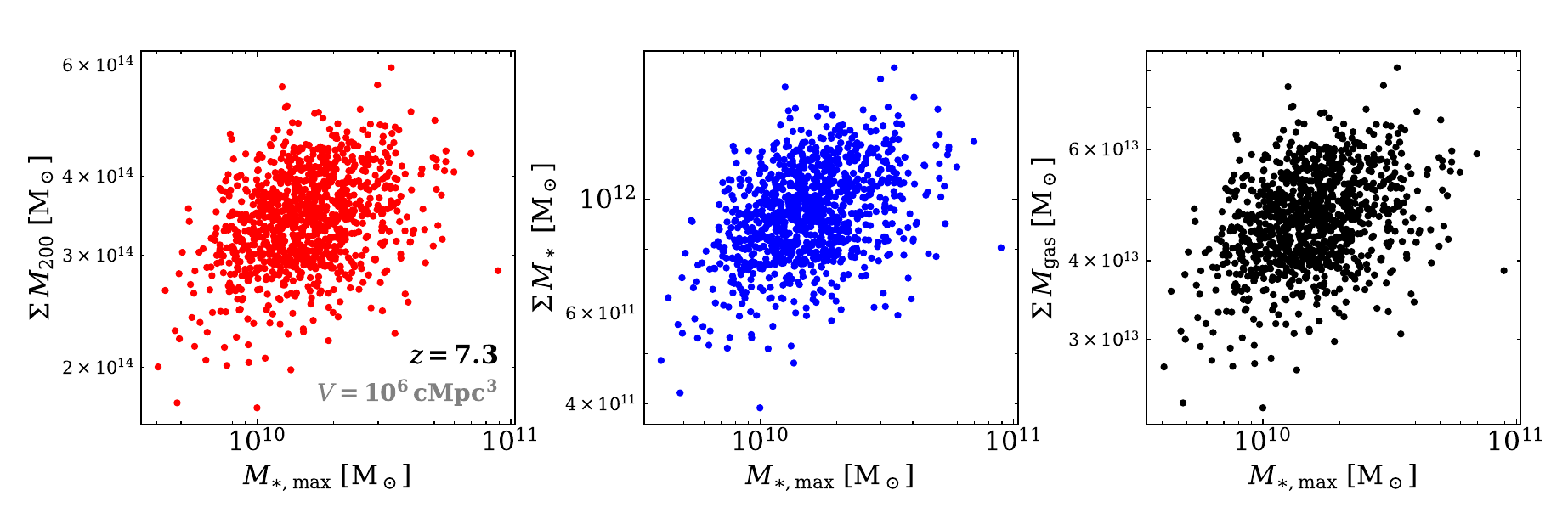}
\vspace{-0.5cm}
\caption{The most massive galaxy, $M_{\rm \ast, max}$, and the sum of the total (including dark matter; left), stellar (middle), and gas (right) mass enclosed for 1,000 subboxes of (100\,cMpc)$^3$ in volume from the FLAMINGO simulation (L1\_m8) at $z\,{\simeq}\,7$. The quantities show a statistically significant positive correlation (Spearman's rank correlation coefficients between 0.32 and 0.37, with $p$-values ${<}\,10^{-25}$), yet the boxes with the highest-$M_{\rm \ast, max}$ galaxies are not necessarily those with the greatest mass sums.}
\label{fig_Msmax_Msums}
\end{figure*}

Figure~\ref{fig_Msmax_Msums} explores the correlation between $M_{\rm \ast, max}$ and the total masses of haloes (left), stars (middle) and gas (right) within each subbox. While here the total mass includes that of the $M_{\rm \ast, max}$ galaxy, we do not find significant changes when it is excluded. As can be seen, there are strong positive correlations (Spearman's rank correlation coefficients between 0.32 and 0.37, with $p$-values ${<}\,10^{-25}$) between $M_{\rm \ast, max}$ and all three types of mass. However, the greatest $M_{\rm \ast, max}$ galaxies are not necessarily embedded in the boxes with the highest total mass. Similarly, we also checked their correlations with the mass functions as well as the stellar and baryon fractions, either averaged over the volume or over the haloes hosting the $M_{\rm \ast, max}$ galaxies, but no strong correlation was found. It is interesting that the greatest-$M_{\rm \ast, max}$ galaxies do not present any strongly enhanced $f_{\ast}$ or $f_{\rm b}$. They are found to have a stellar fraction similar to that of the average $M_{\rm \ast, max}$ galaxies. No correlation with the mass functions (Spearman's $p\,{\simeq}\,0.55$), on the other hand, is consistent with the finding in Sect.~\ref{ssec_SMHM}, where we demonstrated that regions of high clustering are not necessarily places of more efficient galaxy formation. However, we find a significant correlation (Spearman's $p\,{\simeq}\,0.024$) when only the clustering of massive objects (e.g. the HMFs at $M_{200}\,{\gtrsim}\,5\times10^{11}\,{\rm M}_\odot$) is considered. This suggests that the highest-$M_{\rm \ast, max}$ ``monster'' galaxies may form via mergers of other massive galaxies that are abundant in their local volume. We test this idea in the subsequent section. 

\subsection{On the physical origin of the conformity}

\begin{figure}
\includegraphics[width=0.98\linewidth]{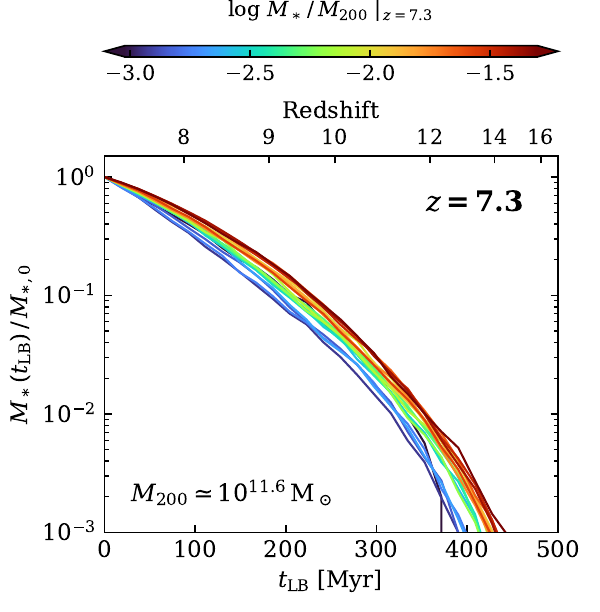}
\vspace{-0.6cm}
\caption{Normalized stellar-mass growth histories of galaxies at $z\,{\simeq}\,7.3$ with $M_{\rm 200}\,{\simeq}\,10^{11.6}\,\mathrm{M}_\odot$ in the FLAMINGO simulation (L1\_m8). Galaxies are divided into 20 bins of stellar-to-halo mass ratio (SMHMR), each with an equal number of systems, and the median history of each bin is shown (colour-coded). Growth is expressed as the cumulative initial mass of stellar particles formed in all progenitors prior to a given lookback time, $t_{\rm LB}$, normalized to that at $z\,{\simeq}\,7.3$. The results demonstrate that galaxies with higher stellar mass at fixed halo mass formed earlier, indicating systematically older stellar populations.}
\label{fig_SFH}
\end{figure}

To explore the physical origin of the conformity, we investigated correlations with star formation properties. We find a strong positive correlation between SMHMR and ${\rm SFR}\,/\,M_{\rm gas}$, but a moderate anti-correlation with sSFR (the plots for these correlations are not shown). The former indicates that massive galaxies with enhanced SMHMR are those experiencing boosted star formation for their gas supply. The anti-correlation with sSFR, however, means that despite their high current efficiency, these galaxies have accumulated a large stellar mass, implying an earlier formation time and older age. This is demonstrated in Fig.~\ref{fig_SFH}, which presents the normalized past mass growth history of galaxies at $z\,{\simeq}\,7.3$ in haloes of $M_{\rm 200}\,{\simeq}\,10^{11.6}\,{\rm M}_\odot$, ranked by SMHMR. Galaxies are divided into 20 bins of equal number, with the median growth history shown for each. The cumulative history was obtained by summing the initial mass of stellar particles formed before a given time in all progenitors, normalized by the mass at $z\,{\simeq}\,7.3$. The result unequivocally shows that massive galaxies with higher stellar mass at a given halo mass formed earlier, as also found by other studies at lower redshifts \citep[e.g.,][]{Lim2016, Matthee2017}. For example, $t_{50}$---the time to assemble half the current stellar mass---is up to 50\,Myr earlier for the highest-SMHMR galaxies. The absolute difference in the formation time is modest, but the systematic nature of this trend is statistically significant. Although limited by the simulation resolution, the highest-SMHMR galaxies also formed their first stars earliest. Combined with our earlier finding of spatial coherence in SMHMR, the earlier assembly of high-SMHMR galaxies indicates that the large-scale conformity arises from accelerated formation of massive galaxies throughout the volume, likely due to their shared initial conditions. 

\subsection{Further implications for observations}

Our results have shown a series of important implications for observations of early massive galaxies, from which a coherent picture appears that there exists a strong large-scale conformity in their evolution out to ${\simeq}$\,100\,cMpc. The detectable signal, however, is significantly weakened for ${\gtrsim}\,50$\,cMpc when galaxies outside the survey footprint are included. Such coherence is a second-order effect, separate from the primary cosmic variance driven by large-scale mass fluctuations. While rooted in the primordial density field, it manifests specifically as a correlation in baryonic properties at fixed halo mass, which we identify as a coherence in star-formation history. This emphasizes the necessity for the full range of large-scale variance to be taken into account, in order to obtain an unbiased and comprehensive understanding of the population. 

Additionally, as mentioned earlier, most studies have assumed the average SMHMR derived from theoretical models or observations, to infer the total mass from their stellar mass. However, as shown in Sect.~\ref{ssec_SMHM}, the SMHMR can be highly dependent on environment. Using the average SMHMR for a highly clustered region may lead to an over-estimation of the total mass by up to a factor of 3 at $z\,{\simeq}\,7$, according to our results. This could mitigate the tension (if any) of the early massive galaxies with LCDM or the galaxy formation model significantly. 

FLAMINGO predicts that the average SMHMR or stellar fraction of the most massive galaxies is much greater (about 4 times; see Sect.~\ref{ssec_Mmaxs}) than usually assumed in the literature. Not only that, it also changes with redshift in a non-trivial way (Fig.~\ref{fig_fs_fb_z}), unlike studies normally adopting the relations from lower-redshift observations. As discussed earlier in this section, this evolution is a composite effect, driven by both the growth in the halo mass of these galaxies and an evolution of the relation at fixed halo mass. 

A striking result from our analysis is that remarkable information can be inferred from a single galaxy, the most massive galaxy, about the entire ${\sim}\,(100\,\mathrm{cMpc})^3$ survey volume. A $M_{\rm \ast, max}$ galaxy presents strong correlations with the evolution and properties of other galaxies in the same volume, as we demonstrated in Sect.~\ref{ssec_SMHM}.The conformity signal remains strong and at the same level for ${\lesssim}\,$50\,cMpc even when galaxies outside the volume are included.

\section[summary]{Summary}
\label{sec_summary}

In this work, we quantified the impact of cosmic variance (CV) on the properties and evolution of early massive galaxies using the large-volume FLAMINGO cosmological hydrodynamical simulation (L1\_m8). Our primary goal was to determine whether large-scale environmental effects, beyond simple Poisson noise and matter clustering, affect comparisons of the abundance and stellar mass of massive galaxies observed at high redshifts ($z\,{\gtrsim}\,5$--$7$) by JWST.

We constructed our analysis by dividing the FLAMINGO simulation box of ($1\,\mathrm{cGpc})^3$ into 1,000 unique sub-volumes of $(100\,\mathrm{cMpc})^3$, matching the typical scale of JWST deep surveys. For each sub-volume, we computed the number counts of haloes and galaxies, the highest halo mass ($M_{\mathrm{h,max}}$) and the highest stellar mass ($M_{\ast,\mathrm{max}}$), and the stellar-to-halo mass relation (SMHMR). We compared the total variance in these quantities to the expectation from a Poisson distribution, where any clustering is random.

Our analysis reveals that the total variance in the number counts of massive haloes ($M_{200}\,{\simeq}\,10^{11.5}\,\mathrm{M}_{\odot}$) at $z\,{\simeq}\,6$ is a factor of 2--3 greater than the Poissonian expectation. Furthermore, the variance in the mass of the most massive halo expected within a JWST-like survey volume is about twice as large as the Poisson prediction at $z\,{\gtrsim}\,4$. This confirms that CV is a significant source of uncertainty when interpreting the abundance and properties of rare, high-redshift objects.

A key result of this work is the discovery of a large-scale conformity in the baryonic assembly of massive galaxies. We find that the star-formation efficiency, as encoded in the SMHMR, is coherently modulated across scales of up to ${\simeq}\,100\,\mathrm{cMpc}$ within a given survey volume. In particular, the stellar mass of the most massive galaxy in a field, $M_{\ast,\mathrm{max}}$, emerges as a sensitive tracer of this coherence: sub-volumes hosting a higher-$M_{\ast,\mathrm{max}}$ galaxy exhibit systematically elevated SMHMR values throughout the entire region. Although this signal weakens when galaxies from neighbouring volumes are included, it nevertheless remains detectable out to ${\simeq}\,100\,\mathrm{cMpc}$, indicating that the conformity reflects a genuine, large-scale physical correlation rather than a purely local effect tied to halo mass or density. This volume-wide coherence underscores the importance of considering both cosmic variance and survey footprint when interpreting early galaxy populations in current and future JWST fields.

This observed conformity is a footprint-dependent manifestation of a real physical phenomenon. Each massive galaxy ($M_{\ast,\max}$) sits at the center of a coherence volume, a region whose constituent galaxies share a common origin in the initial density field and subsequent correlated assembly history. A galaxy is identified as the most massive precisely when a survey footprint successfully captures its dominant coherence volume. Galaxies within this volume show strong mutual correlations, while those outside belong to other such volumes, explaining why the detectable signal is footprint-dependent and is diluted when uncorrelated galaxies are included. This physical interpretation is robustly supported by null tests, where randomized Poisson samples show a complete absence of the signal, ruling out a stochastic origin and confirming that the coherence is imprinted by physical mechanisms correlating galaxy formation across tens of cMpc.

We also characterized the stellar mass fraction $f_\ast= M_\ast/(M_{200}f_{\mathrm{b,cosmic}})$ of these most-massive galaxies in $(100\,\mathrm{cMpc})^3$ volumes. FLAMINGO predicts that their $f_{\ast}$ depends strongly on redshift, rising from a low average value of ${\simeq}\,0.05$ at $z\,{\simeq}\,10$ to a peak of ${\simeq}\,0.2$ by $z\,{\simeq}\,5$, before declining towards $z\,{=}\,0$. This evolution is both more rapid and peaks at a higher value than commonly assumed in models that use a simple log-normal scatter, for example. The distribution of $f_{\ast}$ is also narrower than often adopted. The evolution from $z\,{\simeq}\,10$ to 5 can be naturally explained by a combination of continuous gas accretion and conversion of gas into stars, which increases the stellar fraction while maintaining the baryon fraction close to the cosmic value. Towards $z\,{\simeq}\,0$, the impact of AGN gradually becomes significant, ejecting the gas out of a galaxy into and possibly beyond the CGM. This depletes gas available for further star formation, while the halo mass continues to grow via matter accretion.  

In conclusion, our results from the FLAMINGO simulation demonstrate that the most massive galaxy in a JWST-like survey volume is not merely an extreme outlier but a strong indicator of the large-scale galactic environment within that volume. The presence of strong within-survey conformity on scales of $100\,\mathrm{cMpc}$ must be accounted for to avoid biased interpretations of early galaxy observations. This suggests that the properties of a few extreme objects can reflect the broader cosmic environment of their birth, providing a new lens through which to interpret JWST's discoveries. Simulations of volumes ${\gg}\,$100\,cMpc on a side are required to capture the effects of cosmic conformity. Future work, combining large-volume simulations with deep observational surveys, will be crucial to further constrain the mechanisms driving this large-scale coherence.

\section*{ACKNOWLEDGEMENTS}

SL and RM acknowledge support by the Science and Technology Facilities Council (STFC) and by the UKRI Frontier Research grant RISEandFALL. RM also acknowledges funding from a research professorship from the Royal Society. SL also thanks Tiago Costa and Andrea Ferrara for helpful comments and discussion. This work used the DiRAC@Durham facility managed by the Institute for Computational Cosmology on behalf of the STFC DiRAC HPC Facility (\url{www.dirac.ac.uk}). The equipment was funded by BEIS capital funding via STFC capital grants ST/K00042X/1, ST/P002293/1, ST/R002371/1 and ST/S002502/1, Durham University and STFC operations grant ST/R000832/1. DiRAC is part of the National e-Infrastructure. This work is partly funded by research programme Athena 184.034.002 from the Dutch Research Council (NWO).

\section*{DATA AVAILABILITY}

The data underlying this article will be shared on reasonable request to the corresponding author.

\bibliographystyle{mnras}
\bibliography{GalEnv.bib}

\appendix

\section{Fitting the cosmic variance of mass functions}
\label{sec_appA}

\begin{figure}
\includegraphics[width=1.0\linewidth]{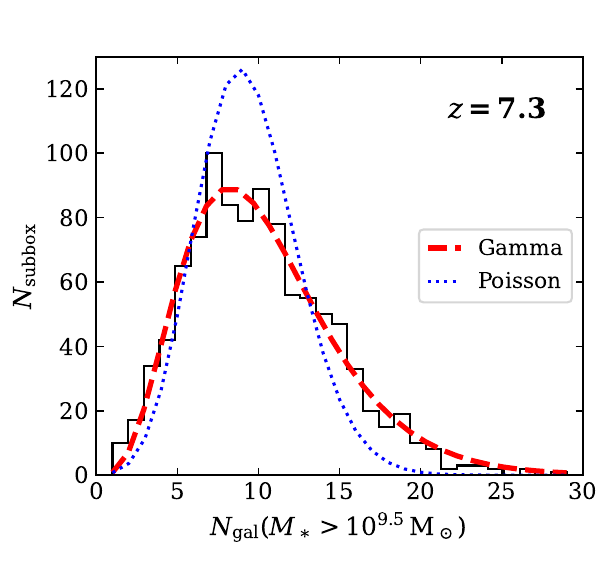}
\vspace{-0.5cm}
\caption{The number count distribution of galaxies with $M_\ast\,{>}\,10^{9.5}\,{\rm M}_\odot$ at $z\,{\simeq}\,7.3$ across 1,000 sub-volumes of (100\,cMpc)$^3$ each, extracted from the FLAMINGO simulation (L1\_m8). The dashed and dotted curves present the best-fit gamma and Poisson distributions, respectively. The distribution is positively skewed, i.e. skewed towards high number count, so that it does not follow the Poisson distribution and instead is well described by the gamma distribution, with a reduced $\chi^2$ of 0.86. The deviation from the Poisson noise demonstrates the strong presence of CV on 100\,cMpc scale. We find the same conclusions for other redshifts and mass thresholds, the best-fitting parameters for which are given in Table~\ref{tab_nM200} and \ref{tab_nMs}. }
\label{fig_fit}
\end{figure}

Figure~\ref{fig_fit} presents a histogram of the galaxy count across 1,000 sub-volumes of (100\,cMpc)$^3$ each, extracted from the FLAMINGO simulation. It shows the number of galaxies with stellar mass greater than $10^{9.5}\,{\rm M}_\odot$ at $z\,{\simeq}\,7.3$ as an example, while we find the same trends hold for other redshifts and mass thresholds. The distribution is not well-described by a Poisson distribution but is instead well-fitted by a gamma distribution, yielding a reduced $\chi^2$ of 0.86. This deviation from Poissonian statistics demonstrates the presence of significant cosmic variance, indicating that the observed variance is driven by underlying large-scale matter distribution, not just simple Poisson counting noise on 100\,cMpc scales. In Tables~\ref{tab_nM200} and \ref{tab_nMs}, we also provide the parameters for the best-fitting gamma distributions from $z\,{\simeq}\,10$ to 0 for several halo and stellar mass cuts. 

\begin{table*}
 \renewcommand{\arraystretch}{1.4} 
 \centering
  \begin{minipage}{150mm}
  \caption{Gamma function fit to the cosmic variance for a (100\,cMpc)$^3$ volume, $n({>}\,M_{200})\,{=}\,x^{k-1}e^{-\frac{x}{\theta}}/\,(\theta^k \Gamma(k))$, from $z\,{\simeq}\,10$ to 0.}
  \begin{tabular}{llccccccc}
\hline
 & & ${\it M}_{200}\,{=}\,10^{11}\,{\rm M}_\odot$ & $10^{11.5}\,{\rm M}_\odot$ & $10^{12}\,{\rm M}_\odot$ & $10^{12.5}\,{\rm M}_\odot$ & $10^{13}\,{\rm M}_\odot$ & $10^{13.5}\,{\rm M}_\odot$ & $10^{14}\,{\rm M}_\odot$  \\ 
\hline
\hline
$z\,{=}\,10.4$ & $k$ & 42$\pm$27 & & & & & & \\
 & $\theta$\textsuperscript{\footnote{The errors for $\theta$'s are negligible, thus not presented.}}
 & 2.5$\times10^{-8}$ & & & & & & \\
$z\,{=}\,7.3$ & $k$ & 15$\pm$0.69 & 4.3$\pm$0.45 & & & & & \\
 & $\theta$ & 1.2$\times10^{-5}$ & 2.5$\times10^{-6}$ & & & & & \\
$z\,{=}\,6$ & $k$ & 23$\pm$1.6 & 11$\pm$0.24 & 1.9$\pm$1.0 & & & & \\
 & $\theta$ & 3.2$\times10^{-5}$ & 6.3$\times10^{-6}$ & 1.6$\times10^{-6}$ & & & & \\
$z\,{=}\,5$ & $k$ & 30$\pm$2.4 & 17$\pm$1.2 & 6.3$\pm$0.31 & 1.5$\pm$0.38 & & & \\
 & $\theta$ & 6.7$\times10^{-5}$ & 1.7$\times10^{-5}$ & 3.9$\times10^{-6}$ & 2.6$\times10^{-8}$ & & & \\
$z\,{=}\,4$ & $k$ & 39$\pm$3.7 & 23$\pm$1.2 & 12$\pm$0.29 & 3.1$\pm$0.28 & & & \\
 & $\theta$ & 1.1$\times10^{-4}$ & 3.7$\times10^{-5}$ & 9.0$\times10^{-6}$ & 2.6$\times10^{-6}$ & & & \\
$z\,{=}\,3$ & $k$ & 49$\pm$6.5 & 32$\pm$3.2 & 19$\pm$1.5 & 8.5$\pm$0.32 & 2.0$\pm$0.67 & & \\
 & $\theta$ & 1.6$\times10^{-4}$ & 6.2$\times10^{-5}$ & 2.0$\times10^{-5}$ & 5.2$\times10^{-6}$ & 1.4$\times10^{-6}$ & & \\
$z\,{=}\,2$ & $k$ & 59$\pm$7.2 & 42$\pm$3.8 & 27$\pm$2.5 & 14$\pm$0.68 & 5.7$\pm$0.19 & 2.4$\pm$1.7 & \\
 & $\theta$ & 1.9$\times10^{-4}$ & 8.2$\times10^{-5}$ & 3.2$\times10^{-5}$ & 1.1$\times10^{-5}$ & 4.1$\times10^{-6}$ & 7.4$\times10^{-6}$ & \\
$z\,{=}\,1$ & $k$ & 71$\pm$4.9 & 51$\pm$6.6 & 35$\pm$2.4 & 20$\pm$1.6 & 10$\pm$0.46 & 4.3$\pm$0.17 & 5.5$\pm$1.6 \\
 & $\theta$ & 1.7$\times10^{-4}$ & 8.7$\times10^{-5}$ & 3.8$\times10^{-5}$ & 1.7$\times10^{-5}$ & 7.3$\times10^{-6}$ & 3.4$\times10^{-6}$ & 2.6$\times10^{-7}$ \\
$z\,{=}\,0$ & $k$ & 54$\pm$4.8 & 45$\pm$3.7 & 35$\pm$3.1 & 23$\pm$0.96 & 14$\pm$0.29 & 7.6$\pm$0.19 & 3.5$\pm$0.39 \\
 & $\theta$ & 1.8$\times10^{-4}$ & 8.6$\times10^{-5}$ & 3.8$\times10^{-5}$ & 1.8$\times10^{-5}$ & 9.6$\times10^{-6}$ & 5.3$\times10^{-6}$ & 2.7$\times10^{-6}$ \\
\hline
\\
\vspace{-15mm}
\end{tabular}
\label{tab_nM200}
\end{minipage}
\vspace{-0.2cm}
\end{table*}

\begin{table*}
 \renewcommand{\arraystretch}{1.4} 
 \centering
  \begin{minipage}{150mm}
  \caption{Gamma function fit to the cosmic variance for a (100\,cMpc)$^3$ volume, $n({>}\,M_\ast)\,{=}\,x^{k-1}e^{-\frac{x}{\theta}}/\,(\theta^k \Gamma(k))$, from $z\,{\simeq}\,10$ to 0.}
  \begin{tabular}{llccccccc}
\hline
 & & ${\it M}_\ast\,{=}\,10^{9}\,{\rm M}_\odot$ & $10^{9.5}\,{\rm M}_\odot$ & $10^{10}\,{\rm M}_\odot$ & $10^{10.5}\,{\rm M}_\odot$ & $10^{11}\,{\rm M}_\odot$ & $10^{11.5}\,{\rm M}_\odot$ & $10^{12}\,{\rm M}_\odot$  \\ 
\hline
\hline
$z\,{=}\,7.3$ & $k$ & 12$\pm$0.54 & 4.4$\pm$0.14 & & & & & \\
 & $\theta$\textsuperscript{\footnote{The errors for $\theta$'s are negligible, thus not presented.}} & 8.4$\times10^{-6}$ & 2.3$\times10^{-6}$ & & & & & \\
$z\,{=}\,6$ & $k$ & 19$\pm$0.40 & 13$\pm$0.19 & 5.9$\pm$0.045 & 4.7$\pm$1.3 & & & \\
 & $\theta$ & 2.5$\times10^{-5}$ & 9.4$\times10^{-6}$ & 4.0$\times10^{-6}$ & 3.8$\times10^{-7}$ & & & \\
$z\,{=}\,5$ & $k$ & 26$\pm$0.49 & 18$\pm$0.46 & 12$\pm$0.040 & 4.7$\pm$0.11 & & & \\
 & $\theta$ & 5.4$\times10^{-5}$ & 2.4$\times10^{-5}$ & 9.5$\times10^{-6}$ & 3.0$\times10^{-8}$ & & & \\
$z\,{=}\,4$ & $k$ & 38$\pm$3.8 & 25$\pm$1.2 & 18$\pm$0.75 & 12$\pm$0.11 & 2.3$\pm$0.35 & & \\
 & $\theta$ & 1.1$\times10^{-4}$ & 4.8$\times10^{-5}$ & 2.3$\times10^{-5}$ & 9.4$\times10^{-6}$ & 1.9$\times10^{-6}$ & & \\
$z\,{=}\,3$ & $k$ & 47$\pm$11 & 38$\pm$4.4 & 29$\pm$0.83 & 17$\pm$0.94 & 6.9$\pm$0.48 & & \\
 & $\theta$ & 1.7$\times10^{-4}$ & 9.3$\times10^{-5}$ & 5.1$\times10^{-5}$ & 2.2$\times10^{-5}$ & 4.0$\times10^{-6}$ & & \\
$z\,{=}\,2$ & $k$ & 63$\pm$8.0 & 48$\pm$8.2 & 41$\pm$7.0 & 28$\pm$2.0 & 13$\pm$0.30 & 2.8$\pm$0.048 & \\
 & $\theta$ & 2.2$\times10^{-4}$ & 1.2$\times10^{-4}$ & 8.5$\times10^{-5}$ & 4.2$\times10^{-5}$ & 1.2$\times10^{-5}$ & 3.5$\times10^{-6}$ & \\
$z\,{=}\,1$ & $k$ & 78$\pm$7.1 & 61$\pm$6.4 & 53$\pm$5.0 & 45$\pm$5.0 & 22$\pm$0.71 & 7.6$\pm$0.17 & 5.3$\pm$0.26 \\
 & $\theta$ & 2.3$\times10^{-4}$ & 1.3$\times10^{-4}$ & 1.0$\times10^{-4}$ & 6.4$\times10^{-5}$ & 2.0$\times10^{-5}$ & 5.5$\times10^{-6}$ & 3.3$\times10^{-7}$ \\
$z\,{=}\,0$ & $k$ & 57$\pm$6.8 & 52$\pm$7.5 & 46$\pm$5.1 & 42$\pm$4.2 & 26$\pm$0.95 & 12$\pm$0.17 & 3.7$\pm$0.013 \\
 & $\theta$ & 2.6$\times10^{-4}$ & 1.6$\times10^{-4}$ & 1.2$\times10^{-4}$ & 7.6$\times10^{-5}$ & 2.5$\times10^{-5}$ & 8.9$\times10^{-6}$ & 2.7$\times10^{-6}$ \\
\hline
\\
\vspace{-15mm}
\end{tabular}
\label{tab_nMs}
\end{minipage}
\vspace{-0.2cm}
\end{table*}

\label{lastpage}

\end{document}